# Nanobolometers for THz Photon Detection

Boris S. Karasik, Andrei V. Sergeev, and Daniel E. Prober

***Abstract*—This article reviews the state of rapidly emerging terahertz hot-electron nanobolometers (nano-HEB), which are currently among of the most sensitive radiation power detectors at submillimeter wavelengths. With the achieved noise equivalent power close to $10^{-19}$ W/Hz$^{1/2}$ and potentially capable of approaching NEP ~ $10^{-20}$ W/Hz$^{1/2}$, nano-HEBs are very important for future space astrophysics platforms with ultralow submillimeter radiation background. The ability of these sensors to detect single low-energy photons opens interesting possibilities for quantum calorimetry in the mid-infrared and even in the far-infrared parts of the electromagnetic spectrum. We discuss the competition in the field of ultrasensitive detectors, the physics and technology of nano-HEBs, recent experimental results, and perspectives for future development.**

## I. INTRODUCTION

THE desire for higher sensitivity detectors has always persisted as astronomical instrumentation evolved and more demanding space applications arose. Detection of faint fluxes of photons is achieved via either registering the average radiation power absorbed in a detector or by counting the average number of photons per second arriving at a detector. The latter scenario is more common for high-energy astronomy (X-ray, gamma-ray) where photons arrive infrequently. However, as the telescopes and space instruments improve, very low radiation backgrounds set the sensitivity limit, so the interest in long-wavelength single-photon detection grows.

The far-infrared (far-IR) including the THz region is a very important spectral range containing about 98% of all the photons existing in the Universe [1]. Because of the high opacity of the Earth's atmosphere, the accessibility of this range from the ground is limited and the role of space borne instruments is very important [2]. The current trend is to pursue moderate resolution spectroscopy ($\nu/\Delta\nu \sim 1000$) of extragalactic lines using nearly zero-emission telescopes (cryogenically cooled primary mirror). This will demand

B. S. Karasik is with the Jet Propulsion Laboratory, California Institute of Technology, Pasadena, CA 91109 USA (phone: 818-393-4438; fax: 818-393-4683; e-mail: boris.s.karasik@ jpl.nasa.gov).

A. V. Sergeev is with the Department of Electrical Engineering, State University of New York at Buffalo, Buffalo, NY 14260 USA (e-mail: asergeev@eng.buffalo.edu).

D. E. Prober is with the Departments of Applied Physics and Physics at Yale University, New Haven, CT 06520-8284 (e-mail: daniel.prober@yale.edu)

detectors with exceptionally low Noise Equivalent Power (*NEP*), in the range $10^{-19}$-$10^{-20}$ W/Hz$^{1/2}$ [3-9] which is significantly smaller than that of the detectors operating on the recent generation of far-IR space instruments (bolometers with semiconductor NTD-Ge thermometer [10] and Ge-doped photoconductors [11, 12]; their best *NEP* = $10^{-17}$-$10^{18}$ W/Hz$^{1/2}$).

In this paper we overview the state of one of the most promising detector approaches for achieving the new sensitivity goals for THz astrophysics -- the superconducting hot-electron nanobolometer (nano-HEB) [13, 14]. Bolometers relying on the weak thermal coupling between the absorber and the heat sink have been highly successful in sensitive far-IR instruments. They contributed to the accurate determination of the temperature of the Cosmic Microwave Background (CMB) [15] and to the detailed study of the anisotropy of the CMB [16]. The nano-HEB relies on the intrinsic weak coupling between the electrons and phonons in a metal absorber. The absorber is made from a superconducting material, which is biased so it is always partially resistive. Detected photons cause the temperature of the electrons to rise and the resistance to increase. This type of thermometer is often referred to as a Transition-Edge Sensor (TES) since it operates within the rather narrow temperature interval where the resistance of a superconducting sensor changes from zero to the normal metal value. In recent years, TES based bolometers played an important role as single-photon quantum calorimeters for X-ray spectroscopy [17] and for the secure quantum communication applications using near-IR single photons [18]. The nano-HEB extends the ability of bolometers to detect both small power and small amounts of energy potentially corresponding to single sub-THz photons.

The paper is organized as follows. First we discuss the far-IR backgrounds in space that eventually set the sensitivity requirements for an ideal detector (Sec. II). The physics and operational principles of the nano-HEB in the power detection and single-photon detection modes, and the model predictions are discussed in Section III. Section IV and V present the results on the power detection and single-photon detection correspondingly. In Section VI we describe other competing approaches to sensitive THz detectors. In Section VII we discuss possible ways for improvement of the nano-HEBs and Section VIII concludes the paper.

## II. BACKGROUNDS AND NEP REQUIREMENTS

In contrast to the heterodyne sensors whose sensitivity is limited by zero-point energy [2, 19] the sensitivity of a direct detector is fundamentally limited only by the fluctuation of the energy in the radiation background seen by the detector. The lowest intensity thermal radiation backgrounds are found in space. In the far-IR and mid-infrared (mid-IR) ranges, which are the focus of this paper, the backgrounds correspond to either the photon remnants of the Big Bang, the Cosmic Microwave Background (CMB) [20] or the interstellar dust emission, the Cosmic Infrared Background (CIB) [21-23]. The combination of these continua defines the irreducible noise associated with the fluctuation of the number of photons impinging an ideal detector with an optical bandwidth $\Delta\nu$.

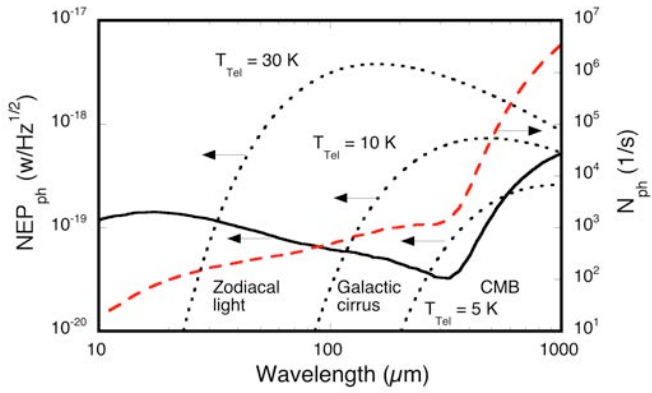

Fig. 1. The *NEP* limited by the background (solid line) and by the telescope emission (5% emissivity, mirror temperature 5K, 10K, and 30K; dots) for an ideal single mode detector operating with a moderate resolution spectrometer ($\nu/\Delta\nu = 1000$) and the rate of photon arrival $N_{ph}$ corresponding to the background (dashes). The latter is less than 1000 s⁻¹ below $\lambda = 300$ μm. The contributions of various radiation continua are calculated from the data of [23].

A detector's Noise Equivalent Power (*NEP*) is a measure of sensitivity and is defined as the radiation power detected with a 1-Hz signal bandwidth with the signal-to-noise ratio (SNR) equal to unity. A general expression for the *NEP* of the background limited photodetector (BLIP) can be found in, for example, [24]:

$$NEP_{ph}^2 = \frac{2}{\eta}\int_{\nu_1}^{\nu_2}P_\nu h\nu\,d\nu + 2\int_{\nu_1}^{\nu_2}P_\nu^2 c^2\,d\nu\Big/mU_\nu\nu^2\,, \qquad (1)$$

where $P_\nu$ is the spectral density of the radiation power incident on the detector, $m$ is the number of polarizations, $\eta$ is the detector quantum efficiency, and $U_\nu = A\Omega\nu^2/c^2$ is the beam throughput ($A$ is the detector effective area, $\Omega$ is the solid angle within which the radiation is coming in). The second term of Eq. 1 is due to the bunching of photons when the number of photons per mode $n_\nu = \left[exp\left(h\nu/k_BT\right)-1\right]^{-1}$ is large. In astronomy, where the backgrounds are low, $n_\nu$ is significant only at millimeter waves [25]. In the infrared and visible, $n_\nu \ll 1$ and the photon noise is given by

$$NEP_{ph}^2 \approx \frac{2}{\eta}\int_{\nu 1}^{\nu 2}P_\nu h\nu\,d\nu = \frac{2}{\eta}P_{rad}h\nu = \frac{2}{\eta}N_{ph}\left(h\nu\right)^2 \qquad (2)$$

assuming that the optical bandwidth $\Delta\nu = \nu_2 - \nu_1$ is narrow. Here $P_{rad}$ is the incident optical power, $N_{ph}$ is the photon arrival rate.

If the radiation background closely resembles the black body radiation spectrum (like, e.g., CMB) and the detector couples to just one radiation mode then $P_\nu = h\nu/\left[exp\left(h\nu/k_BT\right)-1\right]$. For the typical conditions on recent space instruments for CMB studies where $\nu = 100$-500

GHz and $\Delta\nu/\nu \sim 0.3$ [10], the $NEP_{ph}$ is in the range $10^{-18}$-$10^{-17}$ W/Hz$^{1/2}$ for $T = 2.7$ K. In practice, even less sensitive detectors are sufficient compared to this idealized case since the non-zero emission of the relatively warm telescope mirror ($T_{Tel} \sim 40$ K) contributes noticeably to the photon noise.

BLIP operation becomes more demanding when it comes to line spectroscopy with a moderate resolution $\nu/\Delta\nu \sim 1000$ corresponding to a typical width of Doppler-broadened submillimeter lines in extragalactic molecular clouds (see Fig. 1). Several infrared space missions (e.g., IRAS, ISO, Spitzer, Akari) featuring a cryogenically ($\leq 6$ K) cooled primary mirror for elimination of the thermal emission have been flown in past three decades. Several new far-IR mission concepts have been studied in recent years [5-9]. They all considered a far-IR moderate resolution spectrometer as one of the key instruments. A relatively near term realization of such a telescope will be the Japanese Space Infrared Telescope for Cosmology and Astrophysics (SPICA) with potentially two far-IR spectral instruments: an imaging Fourier Transform Spectrometer (FTS) called SAFARI [4] and a grating spectrometer BLISS [3].

Figure 1 illustrates the effect of the cold telescope on the radiation background seen by a single-mode detector pointed at some dark part of the sky with very low luminosity [22, 23]. The associated photon noise $NEP_{ph}$ is very low ($10^{-19}$-$10^{-20}$ W/Hz$^{1/2}$) throughout the entire far-IR and mid-IR parts of the spectrum. A radiatively cooled telescope ($T_{Tel} \sim 30$ K) has $NEP_{ph}$ that is several orders of magnitude higher above 1 THz than a telescope cooled to 5K. Interestingly, the low background condition affects the way a detector might (or should) operate given the very low arrival rate of the far-IR photons, $N_{ph}$. This rate remains under $\sim 1000$ sec⁻¹ from 1 THz to 30 THz. In this situation, an ideal detector should either have a time constant of the order of 100 ms or longer, or be able to detect individual THz photons with high fidelity. The situation is complicated by the fact that the background conditions vary across the sky so a detector capable of counting single photons under low optical loading may not be able to count the photons corresponding to a much more intense background if the detector maximum speed is insufficient. In this case some gradual transition to the power detection mode would be useful. The nano-HEB can operate in both photon-counting and power detection modes with high power and energy resolution, which makes it an interesting candidate for these future space THz applications. Both regimes of operation are discussed in the following sections.

## III. CONCEPT AND UNDERLYING PHYSICS OF HOT ELECTRON BOLOMETERS

### A. Brief history and outline of operation

Hot-electron bolometers (HEB) have been known for quite a while, since the first work on the InSb hot-electron heterodyne detector [26-28]. After the hot-electron effect was realized in superconducting (metal) film bolometers [29-31] interest in the HEBs grew due to the promise of achieving a low-noise THz heterodyne mixer with a large intermediate frequency (IF) bandwidth and low local oscillator (LO) power, unavailable with other detector technologies in early 90s [32]. Eventually, NbN HEB mixers became widespread and are now employed in many ground based, suborbital, and space instruments (see [19] in this issue for the review).

A millimeter-wave Nb direct detector HEB has been demonstrated for the first time in [33] but did not receive serious attention because of the lack of applications. Later, the NbN HEB detector has been found useful for short laser pulse detection [34, 35]. The extremely short electron-phonon (e-ph) relaxation time $\tau_{e-ph}$ =10-20 ps found only in thin NbN superconducting films [36] makes this material stand out among other superconducting materials, even those with similarly high critical temperature (e.g., in NbC with $T_C \approx 9$ K, $\tau_{e-ph} \approx 1$ ns [37, 38]).

In order to bypass the IF bandwidth limitation in widely used superconducting Nb, a nanobolometer version of the heterodyne detector was proposed [39]. While Nb has a relatively long e-ph time, $\tau_{e-ph} \approx 1$ ns, use of non-superconducting contacts allows fast diffusion outward of thermal excitations (cooling) from the short bolometer (length $l \leq 1$ μm) and thermal response times as short as 20 ps have been achieved for heterodyne mixers [40, 41]. Similar enhancement of the IF bandwidth was also observed in NbC HEBs [37]. This is known as a diffusion-cooled bolometer.

The desire to achieve even faster detector response drove the interest in the HEB based on high-$T_C$ superconductors (YBCO). Indeed, a number of results demonstrating fast detection of visible [42, 43], near-IR [44, 45], mid-IR [46], and far-IR [47] radiation have been reported, though the sensitivity was poor. This drawback was explained by the large Kapitza resistance at the film-substrate interface [48] and the dominating phonon heat capacity $C_p$ in the YBCO devices operating at 70-80 K which never allowed biasing the detector to enhance the response at the time intervals shorter than the phonon escape time $\tau_{es} \approx 1$ ns [49]. A typical value of $C_p/C_e$ ($C_e$ is the electron heat capacity) for YBCO was about $10^2$ whereas opposite ratio $C_e >> C_p$ holds at low (< 4K) temperatures where the phonons are not a bottleneck for the releasing the energy from electrons to the substrate.

Perhaps the first paper appreciating the potential of the HEB for untrasensitive detection was that by Nahum and Martinis [50]. They proposed using a long SNS junction (N = normal metal with no pair tunneling between S-contacts) as a bolometric radiation absorber coupled to an antenna. In the following work [51, 52], experimental studies of such normal metal HEBs have been carried out proving the feasibility of low thermal energy fluctuation (TEF, or phonon noise) $NEP_{TEF} \approx 10^{-18}$ W/Hz$^{1/2}$, which is of interest for astrophysics instruments for use in space. A change of the electron temperature $T_e$ is read via a change of the tunnel current in an attached NIS junction. Several approaches to the readout have been discussed as well as the possibility of blocking the electron diffusion by making the device coupling to the antenna through a capacitor [53].

The superconducting HEB appears to offer better performance than the SNS HEB with its normal–metal absorber. The NIS tunnel junction readout of the normal-metal absorber has high impedance. Thus, it is hard to instrument with very low noise, and is difficult to form an array. The SQUID readout matches well to the superconducting HEB, and is compatible with producing a cryogenic multiplexed array. Moreover the SQUID readout has very low noise itself and also provides negative electro-thermal feedback (ETF), thus speeding up the thermal response [54].

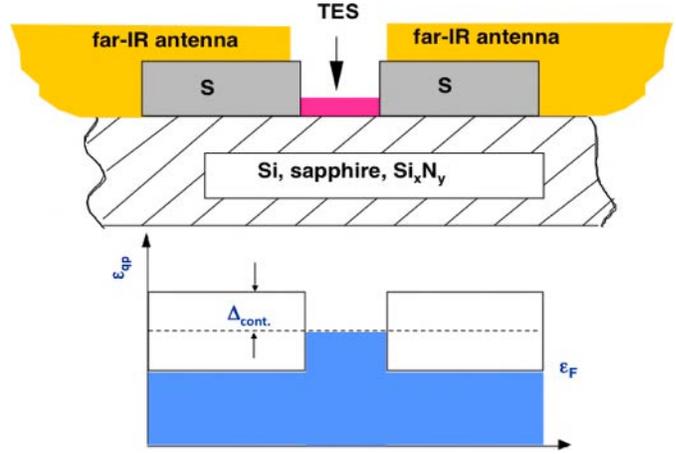

Fig. 2. Top: A crossectional view of the nano-HEB detector. A low-$T_C$ TES device is fabricated on a solid dielectric substrate between superconducting contacts (S) preventing diffusion of the thermal energy. Far-IR radiation couples to the TES through a planar antenna or a waveguide. Bottom: The energy diagram showing the energy gap in the contacts that enables Andreev reflection. The energy gap in the TES is nearly zero since it operates in the resistive state.

Currently, the main interest for the direct detector HEB is in single-photon detection applications. In the visible and near-IR ranges, a ~ 10 μm × 10 μm HEB device can simultaneously be a good absorber of radiation with $\lambda \leq 1.5$ μm and still have a small enough size that the total energy fluctuation due to the detector noise $\delta E << hc/\lambda$. This has been realized in a HEB TES based on W [18, 55, 56], Ti [57, 58], and Ir [59], and with some S-N bilayer materials [60, 61] . Such detectors can be useful for statistical discrimination of the number of simultaneously absorbed photons, which is important for realization of some Quantum Key Distribution (QKD) protocols in schemes for secure quantum communication using single photons. Another application is single photon calorimetry for astrophysics where the small $\delta E$ allows for resolving a multicolor spectrum of the incident flux of photons without using any dispersive optics in front of the detector [62]. A recent Workshop on Single Photon Counting Detectors posted a large number of presentations featuring these HEB applications [63].

### B. Concept of the nano-HEB

A review article [64] gives a good overview of the non-equilibrium (including hot-electron) superconducting detectors and is still current in many aspects. We set here a narrower goal to overview only the smallest HEB detector (often submicron) implementations – nano-HEBs – where the ultrasmall $C_e$ and $G_{e-ph}$ may result in record small $NEP_{TEF}$ and $\delta E$. Most of the practically important results have appeared after paper [64] was published.

The geometry of the TES nanobolometer is in Fig. 2. Superconducting HEBs have all the three critical functions of a bolometer – absorber, thermometer, and thermal conduction – integrated into a single element. The general nano-HEB device configuration is similar to what has been used in the SNS HEB [50] or in optical HEB [18, 55]. The role of superconducting contacts (S) is to create thermal barriers by means of Andreev reflection [65]. Compared to the SNS device, the nano-HEB uses a TES as a thermometer, that is, an additional NIS junction is not needed. Compared to the optical (large size) TES the nano-HEB pushes the fabrication limit and also enters the regime where the penetration of superconductivity from the contacts into the bolometer can affect the TES's characteristics [66].

Since the HEB operates in the resistive state it has practically no superconducting gap and hot electrons are surrounded by an energy barrier of the contact $\Delta_{cont}$. Only a tiny fraction of thermal electrons with energy exceeding $\Delta_{cont}$

$$dn/n \propto \int_{\Delta_{cont}}^{\infty} \left[ \exp\left( \varepsilon / k_B T \right) + 1 \right]^{-1} d\varepsilon \qquad (3)$$

can diffuse into the contacts and deposit energy there. The rest ($\varepsilon < \Delta_{cont}$) experience Andreev reflection, that is the process when an electron-like quasiparticle (qp) at the N-S boundary recombines with a hole-like quasiparticle moving in the opposite direction, to form a Cooper pair [65]. The pair carries electrical charge into the contacts whereas the energy flux is equal to zero.

The TES itself absorbs the radiation as a normal metal. Since the sensor's size is much less than the wavelength and the plasma frequency in metals is in the ultraviolet range, the bolometer represents a lumped element resistor with the impedance equal to the TES's normal resistance $R_N$ for the antenna circuit. The antenna away from the Nb contacts can be constructed of thick Nb, NbTiN, Al, or a non-superconductor such as Au. The low frequency leads for readout are connected to the antenna, not the Ti bridge, and are not shown in Fig. 2.

The thermometer function employs the transition edge of the superconducting transition, the strongly temperature-dependent $R(T)$ curve. The change of resistance is used to detect a change in power (bolometer), or an increment of energy from absorption of a single photon (calorimeter). The terms bolometer and calorimeter are often used interchangeably. The superconducting contacts (S) and the antenna are needed because the lateral size of the absorber (HEB) must be very small even for infrared radiation. This is a result of the nature of thermal isolation of electrons in the HEB, which occurs due to the weak internal coupling between the electron and phonon subsystems. The electrons relax their energy via emission of phonons during the period of time $\tau_{e\text{-}ph}$, so an equivalent e-ph thermal conductance $G_{e\text{-}ph}$ can be introduced as $G_{e\text{-}ph} = C_e / \tau_{e\text{-}ph}$. The e-ph relaxation time sharply increases with decreasing of temperature and can be $\sim$ ms at 0.1 K.

The electron motion, diffusion in such films, is roughly like that of the normal metal, as is the heat capacity, $C_e$. This is because there are few superconducting pairs very near $T_C$ and the energy gap is small or zero, so diffusion is not impeded. In this case, the thermal healing length is $L_T \approx (D\tau_{e\text{-}ph})^{1/2} \sim 300$ μm ($D \approx 1$ cm²/sec is the electron diffusivity in thin metal films which are usually disordered). This distance is quite large and weak e-ph relaxation cannot dominate unless the electron diffusion out the ends of the bolometer is prevented. The small device volume $V$ will decrease $G_{e\text{-}ph}$ proportionally as $C_e \sim V$. As the e-ph thermal conductance is made small, the rms contribution of the fundamental noise mechanism – the thermal energy fluctuation – also becomes small:

$$NEP_{TEF} = \sqrt{\xi k_B T_e^2 G_{e\text{-}ph}} , \qquad (4)$$

where $\xi = 4$ if $T_e = T$, and $\xi = 2$ if $T_e >> T$ [67].

The intrinsic detector sensitivity expected for energy is given by the rms energy resolution, $\delta E$. This limiting sensitivity of the detector itself is set by the TEF due to phonon exchange between the detector and bath, in units of energy $\approx k_B T$. $\delta E$ scales as

$$\delta E = \kappa \sqrt{k_B T_e^2 C_e} \qquad (5)$$

with the prefactor $\kappa$ determined by the relative contribution of the Johnson noise and by the strength of the electrothermal feedback (ETF) [68]. The sensitivity of the TES improves as $T_C$ is reduced, but the time constant increases. Here the effect of the SQUID readout is very helpful. The SQUID can be voltage biased in series with the TES thus providing a negative ETF. This negative ETF gives a faster response, and also provides improved energy resolution [69].

### C. Device size constraints

For single photon calorimetry and sensitive power detection in the far-IR and THz range, very small devices are needed, with volumes $\approx 10^{-2}$ μm³, compared to maybe $\geq 1$ μm³ in the NIR. Thermalization processes must be carefully considered in designing the nanobolometer. The TES dimensions and materials and the contact material must be carefully chosen to ensure that the deposited energy thermalizes fully, only in the volume of the superconductor, and that this happens rapidly compared to the energy loss into the substrate via phonons. The lower limit on device length is imposed by two effects. First, the length of the TES should be larger than the coherence length in the normal metal, $L_T = (\hbar D/4\pi k_B T)^{1/2}$, otherwise the difference between the critical temperatures of the microbridge and the superconducting contacts will be washed out by the proximity effect. At $T \approx 100$ mK and the electron diffusivity $D \approx 2$ cm²/s (moderately impure film) this yields $l >> 2L_T = 100$ nm. Second, if the length is too small, the hot quasiparticles generated by absorbed photons with energies greater than $\Delta_{cont}$ can escape from the TES before thermalization. The escape of the non-thermalized electrons would reduce the quantum efficiency of the detector at frequencies $\nu > \Delta_{cont}/h$ and give variability of the response to monoenergetic photons. For $Nb$ $\Delta_{cont}/h$ corresponds to 360 GHz. One way of preventing deterioration of the spectral response at high frequencies is to make the length of the TES greater than the electron diffusion length at the contact energy gap $l_{e\text{-}e} = [D\tau_{e\text{-}e}(\Delta_{cont})]^{1/2}$ ($\tau_{e\text{-}e}$ is the electron-electron inelastic collision time). Taking $\tau_{e-e}(\Delta_{cont}) \approx 7 \cdot 10^{-11}$ s [70] we estimate this length to be of ~ 90 nm. Thus, if $l \sim 1$ μm these difficulties can be safely avoided. Even though the devices as short as 0.5 μm have been achieved [71], optical measurements have been only performed in HEB detectors with $l \geq 1$ μm [72-74]. The data have shown no signs of sensitivity degradation up to ~ 1 THz [72].

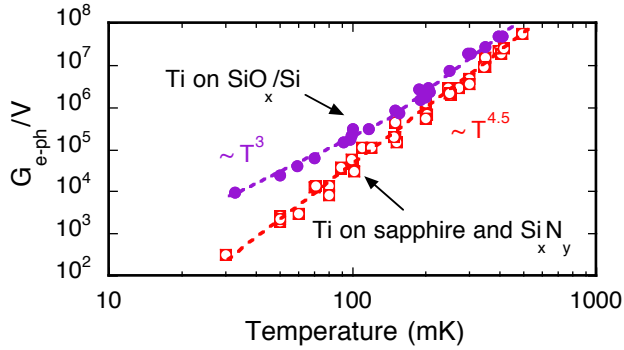

Fig. 3. $G_{e-ph}$ in several 25-40 nm thin large volume Ti films on sapphire and $Si_3N_y$ (squares) and on naturally oxidized Si (circles) [75].

Impure superconducting films can still demonstrate good superconducting properties if the thickness is $\geq 10$ nm. In this case, the bridge width, $w$, is defined by the requirement that $R_N = 50$-100 Ω to match the device with a planar antenna. For many materials this corresponds to a bridge area of roughly one square (ie: 1×1 μm² for 0.1 K). The lateral size can be slightly reduced (by 1.5-2 times) if the detector works at 0.3 K (both $L_T$ and $l_{e-e} \sim T^{-1/2}$).

### D. Thermal conductance

Low thermal conductance $G_{e-ph}$ is the key to achieving high sensitivity in nano-HEBs. Beside the small geometrical size, the device material plays an important role since it defines both the temperature dependence of $G_{e-ph}$ and its prefactor which varies between different materials.

Relatively thin films are required for the HEB. The reason is twofold: a. smaller volume helps to increase the sensitivity b. larger sheet resistance provides a better impedance match between the bolometer and the antenna or waveguide at THz frequencies. In general, thin (thickness $d = 10$-50 nm) superconducting films are fairly disordered. Even if the material itself is pure, collisions of electrons with boundaries increases the resistivity in very thin layers.

The effect of the material on the temperature dependence of the thermal conductance $G_{e-ph}$ comes through the e-ph time $\tau_{e-ph}$, whereas $C_e$ is not usually affected. Since $C_e = \gamma V T$ ($\gamma$ is the Sommerfeld constant), the typical temperature dependence $\tau_{e-ph}(T) \sim T^{-n}$ translates into $G_{e-ph}(T) \sim T^{n+1}$. Table I presents parameters of the available thin film superconducting materials whose $T_C$ is below 0.5 K. Note that S-N bi-layer materials commonly used as TES thermometers are not very suitable for the far-IR HEBs since they cannot be made thin. As a result, both $G_{e-ph}$ and $C_e$ are large, and $R_N$ is small and cannot be impedance matched to an antenna.

A detailed study of the e-ph relaxation in Ti [75] yielded some interesting details for the temperature dependence of $G_{e-}$

$ph$ and of its dependence on the substrate material (see Fig. 3). $G_{e-ph}(T)$ exhibited $n \approx 3.5$ for several Ti films deposited using different techniques. However, in the films grown on naturally oxidized Si $n$ changes from 3.5 above 200 mK to $n = 2$ below 200 mK. This behavior finds a theoretical explanation in the framework of the Pippard's model which takes into account interference of the bulk e-ph scattering and elastic electron scattering on vibrating impurities, defects, and boundaries [79, 80].

In the classic Pippard model [81], effects of disorder on $G_{e-ph}(T)$ are described by a single parameter $q_T\ell$, ($q_T = k_B T / \hbar u$ is the wavevector of a thermal phonon, $u$ is the sound velocity, $\ell$ is the electron mean free path due to scattering from impurities. The e-ph interaction is nonlocal and the characteristic size of the interaction region is of the order of $1/q_T$. In the diffusive limit, $q_T\ell << 1$, a phonon interacts with an electron that diffuses over the interaction region. In this strong "dirty" limit the electron energy relaxation rate is mainly due to the interactions with transverse phonons [81, 82] and it strongly decreases with the disorder:

$$\frac{1}{\tau_{e-ph}} = 9.1 \cdot \frac{3\pi^4 \beta_t}{10} \frac{p_F \ell T^4}{(p_F u_t)^3} \qquad (6)$$

where $\beta_t$ is the coupling constant, $p_F$ is the Fermi momentum, $u_t$ is the transverse sound velocity, and a coefficient 9.1 describes the averaging over all electron states contributing to $G_{e-ph}$ [38]. It turns out that in thin metallic films and nanowires, the transverse phonons also dominate in the opposite, quasi-ballistic case, $q_T\ell >> 1$, in a wide temperature range $T \leq 100$ K, where the e-ph relaxation rate increases with temperature as $T^2$ [83]. For this reason, the dependences of the e-ph relaxation rate (close to $T^3$) widely observed at low temperatures may have nothing to do with the pure e-ph interaction, i.e., with the electron scattering on longitudinal phonons. In fact, such a dependence can originate due to the crossover from the $T^2$-dependence at $q_T\ell >> 1$ to the $T^4$-

TABLE I
PARAMETER OF LOW-$T_C$ MATERIALS FOR HEBs

| Material | $T_C$, mK | Substrate | $\gamma$, J/(K m³) | $G_{e-ph}(T_C)$, pW/K | $n$ | $G_{e-ph}/l$@100mK, W/(K m³) | Reference |
|---|---|---|---|---|---|---|---|
| Ti | 359 | $Si_3N_y$/Si | 310 | 950 | 3 | $3.1×10^5$ | [57] |
| Ti | 351 | $Si_3N_y$/Si | 310 | 13300 | 3 | $8.2×10^5$ | [76] |
| Ti | 430 | Sapphire or $Si_3N_y$ | 310 | varies | 3.4 | $4.0×10^4$ | [75, 77] |
| Ti/Au | 96.4 | $Si_3N_y$/Si | - | 1500 | 3 | $5.4×10^6$ | [76] |
| Ti/Pd | 105 | $Si_3N_y$/Si | - | 8.4 | 3 | $2.1×10^5$ | [60] |
| W | 80 | Si | 136 | 0.94 | 3 | $1.8×10^5$ | [55] |
| W | 95 | Si | 136 | 1200 | 3 | $1.9×10^4$ | [54] |
| Hf | 300 | Sapphire | 160 | $1.0×10^5$ | 4 | $2.4×10^4$ | [77] |
| Ir | 114 | Si | 380 | 12 | - | $5.0×10^4$ | [59] |
| $TiN_x$ | 51 | Si | - | 2.4 | 2.7 | $6.3×10^4$ | [78] |
| $TiN_x$ | 52 | Si | - | 1.2 | 3 | $1.0×10^5$ | [78] |

$n$ is the exponent in the temperature dependence $\tau_{e-ph}(T) \sim T^{-n}$

dependence at $q_T \ell \ll 1$ in the electron relaxation due to scattering on transverse phonons [84]. $G_{e-ph}(T)$ on sapphire or $Si_xN_y$ (see Fig. 3) is an example of such a crossover.

Many experiments with thin films of metallic alloys demonstrated the opposite tendency: the e-ph relaxation rate *increases* due to disorder, and $n = 2$. This obvious contradiction with the Pippard model has been explained in [79, 80]. The destructive interference that leads to the Pippard's condition takes place if all electron scatterers vibrate as host atoms. If vibrations of electron scatterers and host atoms are uncorrelated, the interference results in the enhancement of the e-ph interaction:

$$\frac{1}{\tau_{e-ph}} = 1.6 \cdot b \frac{3\pi^2 \beta_t T^2}{(p_F \ell)(p_F u_t)} \qquad (7)$$

where a coefficient $b$ ($b_{max} = 0.25$) describes the difference in vibration of the electron scatterers and host atoms and a coefficient 1.6 is due the averaging over the electron ensemble. Disorder-enhanced e-ph interaction has been observed in a number of alloys (see [86] for a review) and disordered films (e.g., Nb [87]). In the light of [79, 80], the behavior of $G_{e-ph}$ on a $SiO_x/Si$ substrate (see Fig. 3) may be explained by a combination of the effect given by Eqs. 6 and 7, where the $T^2$-term should be associated with the inelastic electron-boundary scattering at the Ti-substrate interface. While the interference theory [79, 80] provides a qualitative understanding of the effect of the substrate on e-ph relaxation, further research is necessary to reach quantitative agreement [86].

The agreement of $G_{e-ph}/V$ in real nano-HEBs on $SiO_x/Si$ substrates [71, 85] with the data of Fig. 3 is good; typically $G_{e-ph}/V = 10^4 \text{-} 10^5$ W/(K m$^3$) has been observed at 100 mK. From the practical prospective, sapphire or $Si_xN_y$ substrates are preferred since they provide much smaller $G_{e-ph}$ for the same temperature. W and Hf have a significantly lower Sommerfeld constant $\gamma$ than other material; this is advantageous for achieving a small $dE$ in single photon detection. The final choice is a subject of the availability of a reliable technology for fabrication of thin films with desired $T_C$ and resistivity.

## IV. POWER DETECTION MODE

### A. Signal bandwidth and electrical noise

Recent work [85, 88] addressed the electrical noise in nano-HEBs. The usual understanding of the noise in a bolometer is that it consists of two nearly independent components: the TEF noise and the Johnson noise. The full expressions for them can be found in [89, 90]. The TEF noise is large at the bias points where the bolometer approaches the run-away condition: the point where a dissipation of the Joule power $P_J$ cannot be removed by the thermal conductance mechanism. This is usually quantified by the dimensionless ETF loop parameter $L = dP_J/dP_G$ (in the case of HEB, $P_G$ is the cooling power; due to the e-ph interaction in this case) which can be expressed through the parameters of the bias point [89]:

$$L = (dV/dI - R)/(dV/dI + R) \qquad (8)$$

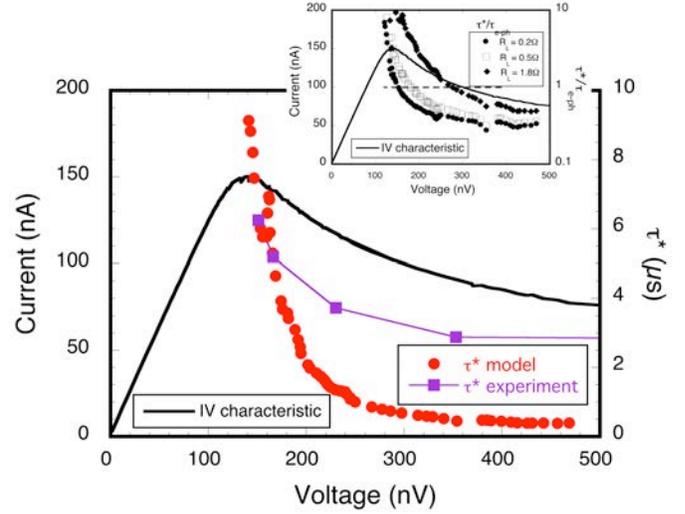

Fig. 4. IV characteristic and time constant for a 6μm × 0.4 μm × 40 nm Ti HEB device on $SiO_x/Si$ substrate at 315 mK [85]. The inset shows the ETF effect on the time constant computed for different load resistors. The dashed line in the inset is a borderline between positive (above) and negative (below) ETF regimes

If the bias circuit provides constant voltage then the current-voltage (IV) characteristic is N-shaped and stable (see Fig. 4; here the large bias range where the device resistance approaches $R_N$ is not shown). From Eq. 8, $L > 1$ if $dV/dI < 0$, and $L < 1$ is $dV/dI > 0$, and $L = 1$ when $dV/dI = \infty$. If the circuit provides constant current (positive ETF for a TES) then the IV curve may became unstable when $L$ approaches unity. Beyond that point the bolometer switches into the normal state irreversibly. Positive ETF is always present to some degree in heterodyne mixer HEBs where the GHz range IF amplifier introduces a large load impedance of 50 Ω. This mode is never used in the low temperature HEBs. The availability of SQUIDs with sufficiently large signal bandwidth and zero input impedance at low frequencies make the nearly ideal voltage bias (negative ETF) always possible.

The ETF affects the bolometer speed so that the actual thermal relaxation occurs with the time constant $\tau^*$ which is different from $\tau_{e-ph}$ [89]:

$$\tau^* = \tau_{e-ph} \frac{dV/dI + R}{2R} \frac{R + R_L}{dV/dI + R_L} \qquad (9)$$

Here $R_L$ is the total resistance of the bias resistor, of the residual resistance in the device, and of the wiring resistance included in series with the TES. For better stability, the goal is to keep $R_L$ as low as possible.

Figure 4 illustrates the dependence of $\tau^*$ on the feedback conditions when $R_L$ still must be taken into account. In this experiment [85], $R_L = 1.8$ Ω is used which is comparable with the device resistance $R$ in some bias points. Compared to $\tau_{e-ph} \approx 4$ μs, $\tau^*$ increases at low biases and decreases at higher biases where the negative ETF becomes stronger ($L >> 1$). A saturation is observed at $\tau^* \approx 3$ μs because of the SQUID bandwidth limit. An inset in Fig. 4 shows what would happen if $R_L$ were smaller than 1.8 Ω. In this case, the transition from the positive ETF ($\tau^*/\tau_{e-ph} > 1$) to the negative ETF ($\tau^*/\tau_{e-ph} < 1$) could occur at smaller bias voltages, that is, mostly a negative ETF would take place everywhere.

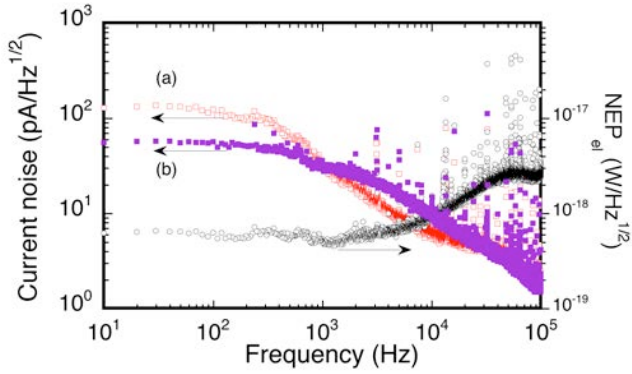

Fig. 5. Output noise spectra and electrical $NEP_{el}$ for a 6 µm × 0.4 µm × 40 nm HEB device at 150 mK [85]. Spectral curves denoted as (a) and (b) correspond to different bias points: a) positive ETF b) no ETF.

This complex behavior for $\tau^*$ reflects in the output noise spectrum as well. The TEF noise roll-off with frequency is at the same cut-off point $(2\pi\tau^*)^{-1}$ as the bolometer responsivity. The Johnson noise by itself is white but the presence of the strong ETF introduces some correlation between the two mechanisms. As a result, the Johnson noise is suppressed wherever $L \gg 1$ (that is, within the bandwidth $\Delta f = (2\pi\tau^*)^{-1}$). It plays little role for the nano-HEBs as the typical SQUID amplifier noise ~ 1 pA/Hz$^{1/2}$ is usually higher.

The magnitude of the TEF noise $i_{TEF}$ is described by the expression $i_{TEF} = NEP_{TEF} \cdot S_I$, where

$$S_I = \frac{\Delta I}{P_{rad}} = \frac{1}{2IR_L} \frac{(dV/dI)/R-1}{(dV/dI)/R_L+1} \frac{1}{\sqrt{1+\left(2\pi f\tau^*\right)^2}} \quad (10)$$

is the detector current responsivity to the incident radiation power $P_{rad}$ [89]. In the strong negative ETF limit, $L \gg 1$, Eq. 10 simplifies and $S_I \approx -1/V$. $i_{TEF}(f)$ rolls off with frequency until it crosses the white part of the noise spectrum (the SQUID noise, that is) (see Fig. 5). The difference between the TEF noise and the typical SQUID noise ($i_{SQUID}$ ~ 1 pA/Hz$^{1/2}$) can be more than tenfold. This results in a large *noise* bandwidth of the detector within which the $NEP$ is nearly constant:

$$\Delta f_n = \Delta f \sqrt{\left(i_{TEF}^2 + i_{SQUID}^2\right)/i_{SQUID}^2} \quad (11)$$

Figure 5 shows an example of the noise spectra measured in different bias points at 150 mK. Here $\tau^*$ ~ 100 µs so the 70 kHz readout bandwidth was sufficient to trace to crossover between the ETF noise and the SQUID noise. Trace (a) corresponds to positive ETF, so $\tau^*/\tau \gg 1$ and the ETF noise cut-off occurs at ~ 300 Hz. Trace (b) corresponds to the situation when ETF does not affect $\tau^*$ so the cut-off of the TEF noise spectrum corresponds to $(2\pi\tau_{e-ph})^{-1}$. In both cases the electrical $NEP_{el}$ defined as

$$NEP_{el}\left(f\right) = \sqrt{i_{TEF}^2\left(f\right) + i_{SQUID}^2}\Big/S_I\left(f\right), \quad (12)$$

remains constant up to ~ 10 kHz where $i_{SQUID}$ begins to dominate.

$NEP_{el}$ usually agrees fairly well with $NEP_{TEF}$. Figure 6

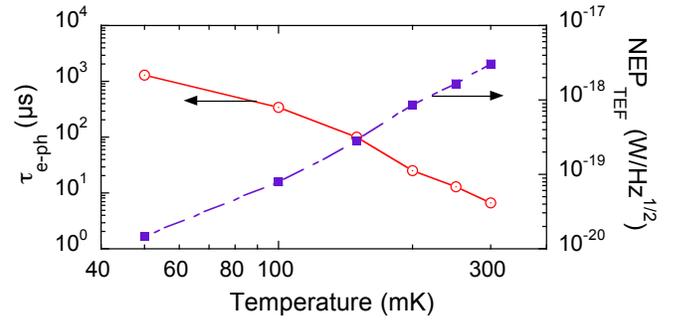

Fig. 6. Physical time constant $\tau_{e-ph}$ and $NEP_{TEF}$ as function of temperature for a 6 µm × 0.4 µm × 40 nm HEB device [85].

shows a summary of the $NEP_{TEF}$ as function of temperature along with the physical time constant measured using a short pulse from a 3-THz quantum cascade laser (QCL) and a very small bias current. Between 150 mK and 300 mK, $\tau_{e-ph}(T) \sim T^4$ which agrees with the theory [91] prediction. The temperature dependence weakens below 100 mK, and correlates with the weakening of the temperature dependence of $G_{e-ph}$ on SiO$_x$/Si (see Fig.3). Nevertheless, the very low $NEP_{TEF}$ = 1.5×10$^{-20}$ W/Hz$^{1/2}$ has been determined at 50 mK. The corresponding $G_{e-ph}$ = 1.6 fW/K, so spurious power as low as 0.01 fW would have a noticeable effect on the electron temperature.

### B. Optical NEP

Optical $NEP$ measurements of THz cryogenic direct detectors are absolutely critical but they are not very common. If the sensitivity is very high then it becomes very difficult to precisely attenuate the calibration signal generated at room temperature without affecting the detector by the high radiation background. The calibration signal carries its own photon noise (see Eq. 2) so its power should be constrained to not only avoid the detector saturation but also to make sure that the photon noise does not exceed the detector noise. For a detector with ultralow $NEP$ = 10$^{-20}$-10$^{-19}$ W/Hz$^{1/2}$, the 1-THz signal with $P_{rad}$ = 0.1-10 aW will already induce significant photon noise. Producing this attowatt level of power is unprecedented and there is no proven recipe on how to build a stable calibration source for this.

Besides our work [73, 74], we are aware of only two reports [92, 93] describing the technique for calibration ultrasensitive detectors with $NEP \le 10^{-18}$ W/Hz$^{1/2}$. All three works use a cryogenic black body radiator with a band defining filter and some procedure for figuring out the amount of $P_{rad}$ impinging on the detector.

In [73, 74], a single mode detector (see Fig.7) was used. The $NEP$ determination procedure consisted of the measurements of bias current $I$ as function of the black body temperature $T_{BB}$ varied in small 0.25 K steps and of the output system noise $i_n$. The HEB was integrated into a 500-800 GHz twin-slot antenna, which is a well known design fabricated using optical lithography. An elliptical Si lens makes the main lobe of the antenna diffraction limited, so only a single radiation mode couples to the HEB device. This makes the calculation of the $P_{rad}$ rather straightforward:

$$P_{rad} = \int_0^\infty \frac{Tr(\nu)h\nu d\nu}{\exp(h\nu/k_B T_{BB})-1} \,. \qquad (13)$$

Here $Tr(\nu)$ is the transmittance of the band limiting filter placed at the mK stage. Since $T_{BB}$ = 1.5-5 K the radiation power follows the Wien's law and the filter must have a very sharp lower frequency edge in order to eliminate any error due to the uncontrolled out of band leakage of the radiation. Free-standing metal mesh filters were used in [73] and a Frequency Selective Surface (FSS) filter [94] was used in [74].

From the initial part of the $I(P_{rad})$ dependence, the detector responsivity $S_I$ was calculated (see Fig. 8). Then the small signal optical $NEP$ was found as $NEP = i_n(P_{rad}=0)/S_I$. One can see that saturation of $\Delta I$ vs $P_{rad}$ naturally occurs much sooner at 50 mK than at 355 mK. The output noise also behaves differently. Whereas $i_n$ at 355 mK monotonically decreases with $P_{rad}$, it exhibits a peak at 50 mK at $P_{rad} \approx 10$ fW. We speculate that the origin of this peak may be in the detection of the fluctuation of power in the impinging radiation. This photon shot noise $NEP_{ph}$ increases with the radiation power. When the detector becomes sensitive enough to detect this fluctuation ($NEP \sim NEP_{phot.}$) the output noise-like signal increases as square root of $P_{rad}$. Eventually, the output saturates, $S_I$ drops, and the output noise decreases. This effect could not be seen at 355 mK since a much greater $P_{rad} \sim 100$ fW would be needed to make $NEP_{phot.}$ large enough to be detected. But at such high power the detector is already saturated.

In this experiment $NEP = 3\times10^{-19}$ W/Hz$^{1/2}$ was measured as the *optical NEP*, which is one of the best numbers found in the literature to date. The quantum efficiency (optical coupling) $\eta$ = 70-80 % was determined from both the ratio of $NEP_{TEF}$ and optical $NEP$ [74] and from the shift of the current-voltage characteristics under optical pumping [73]. Additional improvements are possible by implementing an anti-reflection (AR) coating on the Si lens and by fixing the design of the overlap areas between NbTiN and Ti (see Fig. 7, more details in [73, 74]). In the future, more optical measurements are planned with the described devices especially at higher frequencies (several THz) for which spiral antenna coupled Ti HEBs are already available [73].

## V. PROSPECTS FOR SINGLE-PHOTON DETECTION – MID-IR AND FAR-IR

### A. Minimum detectable energy and energy resolution in HEB

The potential of the HEB for detection of single THz photons can be illustrated using the data of Fig. 6. Indeed, the energy resolution of a bolometer can be roughly estimated as

$$\delta E \approx NEP_{TEF}\sqrt{\tau_{e-ph}} \,. \qquad (14)$$

This is similar to Eq. 5. Figure 9 expresses the data of Fig. 6 in terms of $\delta E$. One can see that $\delta E/h \sim 1$ THz might be possible even for this relatively large HEB device at low operating temperature $\sim$ 50 mK. More detailed analysis of [95] based on the strong ETF limit model [69] gives:

$$\delta E_{rms} \approx 0.3\sqrt{k_B T_C^2 C_e} \,, \qquad (15)$$

For the smallest devices of [71], this would result in $\delta E_{rms}/h =$

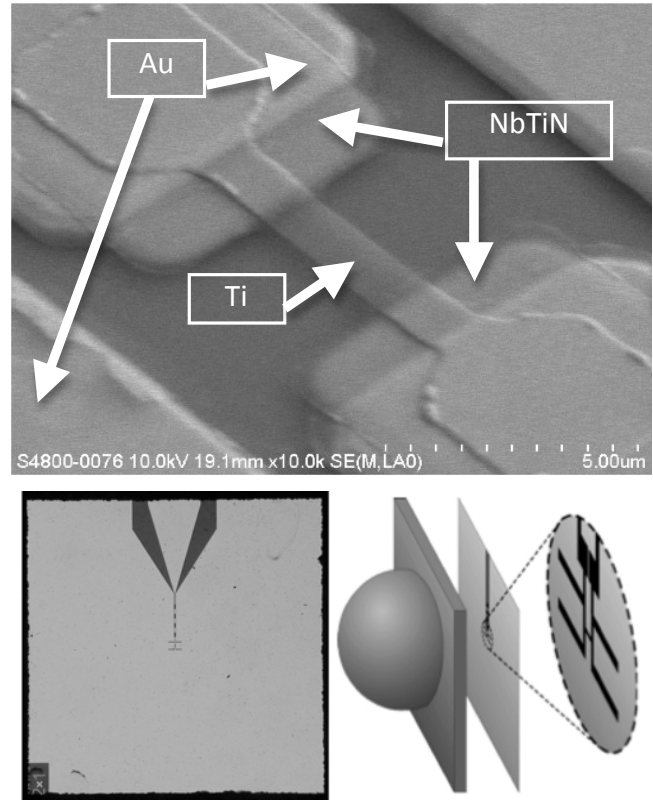

Fig. 7. Top: SEM image of the feeding point of the antenna with a 2μm×1μm Ti HEB device in the center of the coplanar waveguide structure of the planar antenna. Bottom left: a twin-slot antenna chip (4mm × 4mm) with the 500-800 GHz bandwidth. Bottom right: positioning of the chip on a 12-mm elliptical lens.

$50\times T^{3/2}$ MHz/mK$^{3/2}$. If good TES devices of these small sizes were realized one would expect to achieve the desired $NEP \sim 10^{-20}$ W/Hz$^{1/2}$ using photon counting and, possibly, even at 300 mK [95].

Photon counting at THz is needed as the way to achieve better sensitivity for detection of small photon fluxes and/or to relax the ultra low operating temperature requirements. At higher frequencies (mid-IR), however, the HEBs can enable real quantum calorimetry providing useful energy resolution $h\nu/\delta E \sim 10\text{-}100$. This may be an interesting opportunity for future searches for spectral signatures of Earth-like exoplanets using nulling interferometry in the mid-IR (e.g., TPF-I mission concept [96]). When the starlight is suppressed, a distant ($\sim 10$ pc) planet would emit less than a photon per second in the 6-18 μm wavelength range corresponding to important life-tracing molecules [97]. The relatively large speed of the nano-HEB ($\sim 100$ kHz, see discussion of the calorimeter bandwidth in Section VII)) would provide sufficient dynamic range for such observations using photon counting.

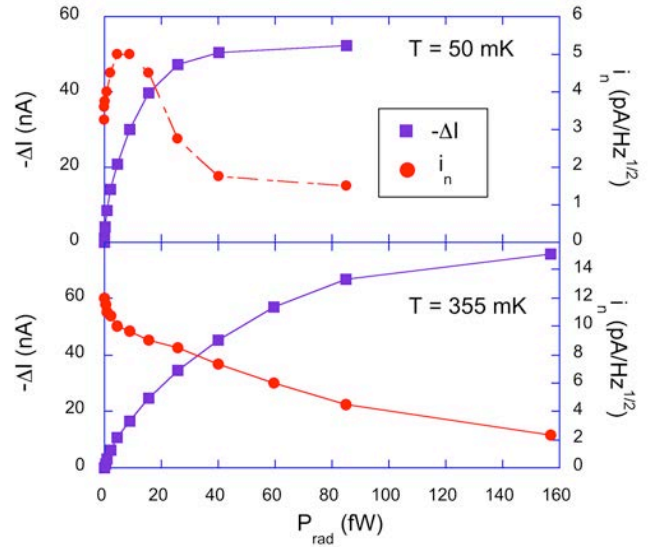

Fig. 8. Experimental data used for determination of optical $NEP$ in a 1 μm × 1 μm × 20 nm HEB device.

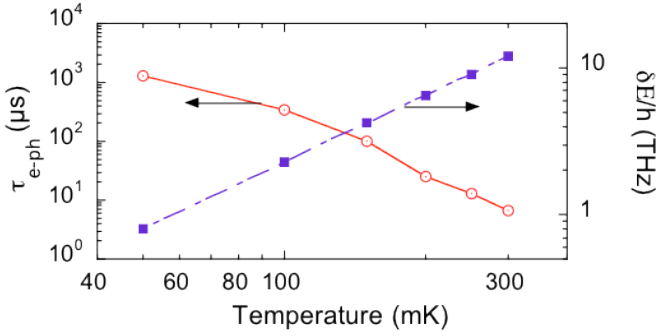

Fig. 9. Physical time constant $\tau_{e\text{-}ph}$ and $\delta E$ as function of temperature re-plotted from Fig. 6

### B. Experiments with "fauxtons"

Detection of single THz photons with HEBs has not been possible so far. However, recent work [98] simulated detection of THz photons using short microwave pulses (20 GHz) of equivalent energy (faux photons, or *fauxtons*). This test method allows for full control of the input amplitude (equivalent to the 'fauxton' energy), as well as precise calibration of the coupling efficiency. It also avoids several issues associated with real optical experiments, including absorption of photons outside the active detector element and the loss of energy by the electron system before reaching a thermal distribution. The fauxton technique thus provides a lower bound on the achievable energy resolution in a real optical experiment.

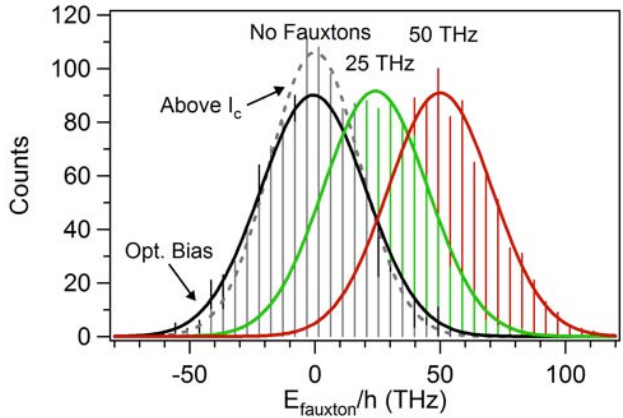

Fig. 10. Histograms of single-shot device response to 50 THz fauxtons, 25 THz fauxtons, and no fauxtons. Response with no fauxtons is measured with the device optimally biased for detection, and also with the device biased above the critical current $I_c$ to determine the amplifier contribution to the measured energy resolution.

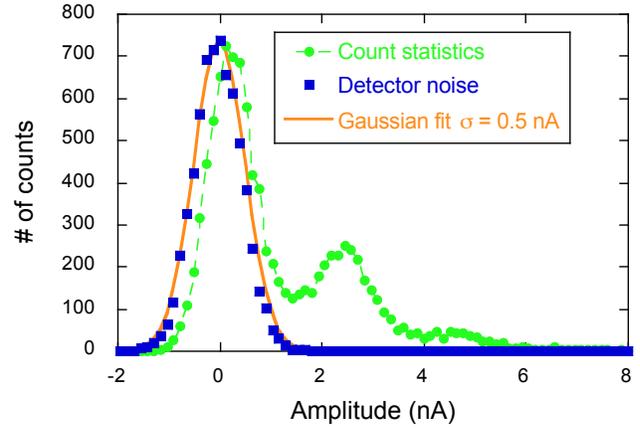

Fig. 11. Count statistics of 8-μm single photons using a 6 μm × 0.4 μm × 20 nm HEB device at 100 mK ($10^4$ events). The amplitudes of photon number peaks follow the Poisson distribution with $\mu = 0.33$.

The device studied had Ti dimensions of 4 μm × 0.35 μm × 70 nm and $T_C = 0.3$ K. The detector response was measured for $10^3$ single-shot pulses for fauxton equivalent frequencies of $E_{fauxton}/h = 50$ THz and 25 THz, as well as with no fauxtons. Figure 10 shows histograms of the number of detected pulses and their heights, for each fauxton frequency. Fitting these histograms to a Gaussian distribution, we found a total energy resolution $\delta E_{FWHM}/h = 2.36 \times \delta E = 49\pm1$ THz. Subtracting the amplifier contribution, an intrinsic detector energy resolution of approximately $\delta E_{FWHM} = 23$ THz was found. This is close to the value predicted, 20 THz, due to ETF noise [98].

### C. Detection of single 8-μm photons

The first detection of real 8-μm single-photons using a 6 μm × 0.4 μm × 20 nm HEB at 50-200 mK was reported in [85]. This was done using a pulsed QCL, which generated short (> 100 ns) optical pulses. Both the detector and the laser were inside a 4-K cryogenic shield in order to avoid any photons arriving from the warm surfaces. The detector was additionally shielded by a millikelvin enclosure. The lightpipe guiding the radiation included some absorbers and filters in order to adjust the pulse energy so the average number of detected photons per pulse was always less than 1. The device did not have any antenna or other means to couple to the radiation efficiently so the optical coupling was very poor.

Figure 11 shows the photon count statistics at 100 mK. The dark counts follow the Gaussian distribution and it has been verified that this is determined by an integral of detector noise over its spectral span. The photon number peaks follow the Poisson distribution $f(n, \mu) \sim \mu^n e^{-\mu}/k!$, where $k$ is the number of simultaneously detected photons, $\mu = 0.33$ is the average number of photons per pulse. The energy interval between photon number peaks is $h\nu$ and the output is fairly linear so the energy resolution of the detector was found to be $h\nu/\delta E_{FWHM} = 1.4$, that is $\delta E_{FWHM}/h = 27$ THz. This resolution remained unchanged from 50 mK to 150 mK and then degrades above ~ 200 mK, completely disappearing near $T_C$.

In the fauxton experiments, a similar $\delta E_{FWHM}$ was found at 300 mK with a similar volume device. Since $\delta E_{FWHM}$ should scale with temperature as $T^{3/2}$ (see Eq. 5), one can say that the energy resolution was noticeably worse with real 8-μm photons. There are several possible causes leading to degradation of $\delta E_{FWHM}$. One of the causes is apparent from Fig. 11 where the center of the peak corresponding to 0 photon number shifts up with laser power. It means that there are some low-energy excitations (either out-of-band photons emitted by the laser, or secondary photons downconverted in the substrate or electrodes, or non-thermal phonons or electrons generated outside the device) contributing to the count statistics [99]. This might also possibly lead to the widening of the photon number peaks. The magnetic field used in this experiment could be another factor degrading the energy resolution by means of widening the superconducting transition and thus nulling the effect of the negative ETF on $\delta E$ [69].

## VI. STATUS OF THE FIELD OF THE SENSITIVE THz DETECTORS

Beside the HEB, several other ultrasensitive detector concepts have been proposed and pursued in the recent decade with the goal to achieve the $NEP \sim 10^{-19}$-$10^{-20}$ W/Hz$^{1/2}$ or single THz photon sensitivity. The motivation has been high but so has been the challenge. To date, the practically achieved $NEP$ in space instruments (e.g., HFI bolometers on ESA's Planck Surveyor mission) is of the order of $10^{-17}$ W/Hz$^{1/2}$ [10]. A two order of magnitude improvement can not be expected to be obtained easily and requires serious study of the unrealized potential of traditional far-IR detectors (bolometers) and the emerging detector approaches.

Significant progress has been made in improving the sensitivity of Si$_x$N$_y$ membrane supported bolometers. The main direction here is to make the suspending leg long and thin in order to reduce the thermal conductance. The recent advance in this technology toward higher sensitivity was a demonstration of a low thermal conductance ($G = 8$ fW/K at 60 mK) in a bolometer suspended on ~ 8 mm long Si$_x$N$_y$ beams [100]. The $NEP_{TEF}$ limit derived from measuring $G$ is $5 \times 10^{-20}$ W/Hz$^{1/2}$. However, successful operation of an array of fully–functioning, optically coupled, SQUID-multiplexed detectors, with uniform electrical and noise characteristics, and meeting all the specifications including ultra–low $NEP$, remains to be demonstrated.

A similar development of a low-G TES for SPICA/SAFARI [92, 101] has resulted recently in the demonstration of an optical $NEP$ below $10^{-18}$ W/Hz$^{1/2}$ [102, 103] The progress has been good and this development effort continues towards meeting the SAFARI's sensitivity goals.

A superconducting kinetic inductance detector (KID) [82, 104, 105], where the high sensitivity is the result of a very low number of quasiparticles $N_{qp}$ (=low generation-recombination noise), potentially can be very sensitive too. Its fundamental noise limit set by the quasiparticle generation-recombination noise is given by $NEP_{GR} = 2\Delta\sqrt{N_{qp}/\tau_{qp}}$ [104] ($\tau_{qp}$ is the quasiparticle life time, $\Delta$ is the energy gap in the detector material). From the physics perspective, this is a similar limit to that set by the TEF noise in bolometers except here the average energy of quasiparticles is $\Delta$, not $k_B T$, and $\Delta >> k_B T$.

In principle, $NEP \sim 10^{-20}$ W/Hz$^{1/2}$ might be possible in an extreme situation when $N_{qp} \sim 100$ and $\tau_{qp} \sim 10$ ms. So far, KID detectors have been developed to the degree of the array demonstration on a ground based telescope [106] with the optical $NEP \approx 10^{-15}$ W/Hz$^{1/2}$. In lab experiments, a much better electrical $NEP \approx 4 \times 10^{-19}$ W/Hz$^{1/2}$ [107] has been derived from the noise measurement. A recent measurement of $N_{qp} \approx 3 \times 10^4$ and $\tau_{qp} \approx 2 \times 10^{-3}$ in an Al KID sample [108] also suggests a similar $NEP_{GR} = 2 \times 10^{-19}$ W/Hz$^{1/2}$ below 180 mK.

The THz superconducting tunnel junction (STJ) concept [109, 110] is based on a superconducting absorber integrated with an STJ and a sensitive RF single-electron transistor, which detects the quasiparticles generated by THz photons in the superconductor. It has not been brought to a practical demonstration yet.

Anther concept using detection of small number of quasiparticles through the change of the charging energy of a small superconducting island has been proposed in [111]. It has demonstrated recently an electrical $NEP = 3 \times 10^{-18}$ W/Hz$^{1/2}$ [112].

An interesting demonstration of the single THz photon sensitive detector has been made in [113, 114] (see [115] and references therein for an overview). The sensors were the quantum-dot structures with the single-electron transistor (SET) read-out. In [113], the absorption of a photon by the quantum dot initiates an electron transition across the magnetic-field-induced energy gap. This transition removes the Coulomb blockade for the other electron states that results in an electron current of $10^6$-$10^{12}$ electrons flowing through the quantum dot. At $T = 50$ mK, an effective $NEP \sim 10^{-22}$ W/Hz$^{1/2}$ was estimated from the dark count rate. Unfortunately, this sensor requires a very strong magnetic fields, ~ 4 T, and, therefore, that makes its use in space challenging. The same research group realized a double dot submillimeter detector without a magnetic field [114]. However, in that structure the electronic noise increases significantly, and the minimum achieved $NEP$ was of the order of $10^{-17}$ W/Hz$^{1/2}$ at 70 mK. The work in this direction continues [116]. A particular interesting practical outcome has been the THz imaging of very cold samples using this photon counting device [117]. Relatively low radiation coupling efficiency (~ few percent) in this promising detector is one of the factors which needs improvement.

Finally, a very recent development [93] concerns a Pb$_{1-x}$Sn$_x$Te(In) based photoconductive detector operating at 1.5 K. Preliminary measurements indicated the $NEP$ as low as $6 \times 10^{-20}$ W/Hz$^{1/2}$ at $\lambda = 350$ μm. Given the very early development phase of this detector, none of the issues related to detector array fabrication, detector radiation coupling, and readout have been addressed yet.

## VII. ISSUES AND FUTURE DIRECTIONS

There are several current issues, which may lead to an immediate improvement of the nano-HEB detector characteristics. The most important is to achieve controllable tuning of the device critical temperature to a desired low value. This problem is common for all TESes; it becomes very difficult when it comes to large TES arrays where all the pixels must be tuned to more or less the same $T_C$ value. A recognized approach for TES thermometers is to use S-N bilayers, but as we mentioned above, this is not a good approach for far-IR HEBs where the absorber is integrated with a thermometer into one element. The unfortunate circumstance here is that elementary superconductors (e.g, Hf, Ir, W), which have sufficiently low $T_C$ in bulk tend to increase it when prepared as thin films. Ti tends to decrease $T_C$ when the film is made thinner but at some point it loses superconductivity abruptly.

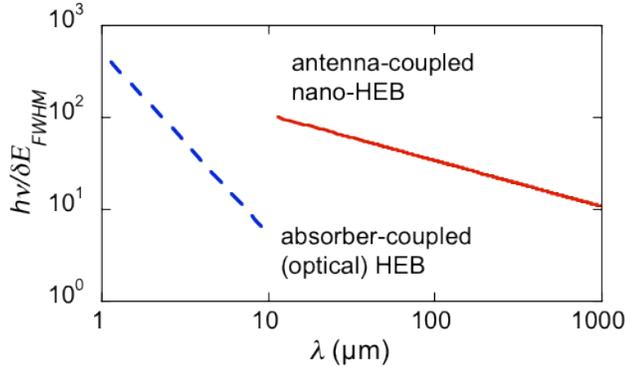

Fig. 12. Expected energy resolution from the visible to the far-IR based on the available approaches to electromagnetic coupling. For antenna-coupled HEBs the temperature increase due to the absorbed photon was assumed to be equal to the superconducting transition width. For optical HEBs the absorber's area was assumed to be 5-times of the wavelength.

Two possible solutions are seen at the moment. One of them is to use films of TiN$_x$ [107] in which $T_C$ can be tuned in a very broad range (<50 mK to 5 K) by the composition of Ti and N$_2$. The films with $T_C \approx 50$ mK [78] are sufficiently resistive for the impedance match to antennas and have low $G_{e-ph}$ similar to other elementary superconductors (see Table I). Another approach is implantation of $^{55}$Mn$^+$ ions into finished devices or into bare Ti films. This has been demonstrated to be a reasonably well controlled process [118]. A similar techniques has been also used successfully for tuning $T_C$ in W using $^{56}$Fe$^+$ ions [119]. One more option is to synthesize some binary alloys (like, e.g. AlMn [120]).

The low $T_C$ issue is the only real challenge at the moment to achieve the fully functioning nano-HEB detector with $NEP < 10^{-19}$ W/Hz$^{1/2}$ suitable for large far-IR arrays. Multiplexed readout was not discussed in this review but a feasible approach using microwave multiplexed SQUIDs is compatible with nano-HEBs [121] which gives no added noise and a large signal bandwidth ~ few 100s kHz.

Getting good energy resolution is another challenge. The fauxton data show that the devices by themselves have $\delta E$ close to the theoretical prediction. Realization of this energy resolution at THz requires good quality (large $I_C$) submicron (small) devices (~ 1 μm × 0.3 μm in-plane size at least), which have not been available yet. The micron-size devices, which are of good quality already, would be useful for mid-IR calorimetry providing the effect of the interaction of radiation directly with the substrate is reduced. Highly efficient spiral antennas are available up to 5-6 THz [122, 123]. Also, mid-IR ($\lambda \sim 10$ μm) antennas have been developed with some success over the past two decades [124-126]. The antenna issue is important since it actually hampers the improvement of the energy resolution in optical TES HEB. Antennas in the NIR are not yet well developed [125].

In the far-IR, the best way to optimize the calorimeter is to adjust its size (electron heat capacity) so the temperature rise caused by a photon drives the device electron temperature through its entire dynamic range that is the superconducting transition width $\delta T_C$. Using Eq. 15 and also assuming $T = 50$ mK, $\delta T_C = 10$ mK, energy resolution $h\nu/\delta E_{FWHM} \sim 100$ can be expected for an HEB device made from a 20-nm thick Ti film (see Fig. 12). The corresponding optimal device in-plane size is 2.5 μm × 2.5 μm even at $\lambda = 10$ μm, that is still smaller than the wavelength. The resolution increases at shorter wavelength as $\lambda^{-1/2}$ but at some wavelength (around 10 μm) one needs to switch to the absorber coupling since the present broadband antennas become too lossy. The absorber area $A$ then needs to be significantly greater than the wavelength (we assumed $A = 25\lambda^2$) and the energy resolution thus degrades. It improves again in the visible (as $\lambda^{-2}$) due to the smaller TES area one can use for visible photons, but the "antenna gap" in the mid-IR is a serious issue.

Another aspect in which the nano-HEB calorimeter differs from, e.g., the X-ray detector is the required response time. In the far-IR, the desired maximum count rate is much larger than the expected photon background, which is $\approx$ 100-1000 s$^{-1}$ (see Fig. 1), so that spectral lines that have much larger count rates can be recorded. Optimum energy resolution, or lower $NEP$, is achieved by cold operation, at, e.g., 50 mK. Unfortunately, at this temperature the thermal response time with only phonon cooling may be too long (~ ms) even with the benefit of electrothermal feedback. We here outline a method to achieve the desired large count rates at 50 mK. We consider an additional cooling 'path', emission of thermal *photons* [127]. These photons are familiar as the Johnson noise of a resistor and can carry energy away from the hot electrons with $T_e > T$. Since $k_B T/h \approx 1$ GHz, the cooling power within the entire spectrum due to this mechanism is $P_G \approx (k_B T)^2/h$. The corresponding *electron-photon* thermal conductance $G_{e-g} = dP_G/dT = 2k_B^2 T/h \approx 30$ fW/K. This compares to the e-ph thermal conductance of the smallest HEB [71] $G_{e-ph} = 0.1$-1 fW/K, and thus *photon* cooling gives a speedup factor up to two orders of magnitude compared to *phonon* cooling alone. This concept can indeed be instrumented, with careful rf design. Using photon cooling will bring the nano-HEB signal bandwidth to ~ 100 kHz at 50 mK. This is very similar to what the current phonon-cooled devices have at ~ 300 mK. Such a bandwidth is common in SQUID amplifiers and has been demonstrated also in the microwave SQUID readout [121].

## VIII. CONCLUSIONS

In conclusion, the hot-electron nanobolometers have reached some maturity, which allow them to be considered as the detectors of choice for the most sensitive applications in the far-IR (e.g., moderate resolution spectroscopy in space).

This field is narrowly defined but the scientific pay-off and the interest are great. There are two missions in development requiring high-sensitivity detectors (SPICA and Millimetron) and the nano-HEB is already meeting the *NEP* requirements for one of the instruments planned (SAFARI). In the longer run, nano-HEBs can enable single-photon detection and calorimetry in the mid-IR and far-IR spectral ranges, which are unprecedented applications. Whereas the far-IR applications are mostly limited to astrophysics, the mid-IR calorimetry may be interesting for single-molecule spectroscopy and free space quantum communication [128].

Beside the advance in fabrication techniques and material synthesis required for nano-HEBs, any breakthrough in optical antennas or in any other methods allowing for efficient coupling of optical radiation to a subwavelength detector would have an immediate impact on the application of HEBs at the short-wavelength mid-IR or NIR ranges. Since the radiation background at $\lambda < 2$ μm is very low, many interesting application ideas may emerge if a practical nano-HEB calorimeter becomes available.

## ACKNOWLEDGMENT

Many colleagues have made critical contributions to the establishment the nano-HEBs through previous and current collaborations. We especially acknowledge Robin Cantor (STAR Cryoelectronics), Michael Gershenson (Rutgers University), David Olaya (National Institute of Standards and Technology), Sergey Pereverzev (Lawrence Livermore National Laboratory), Daniel Santavicca (Yale University), and Jian Wei (Peking University). We also thank our JPL colleagues Bruce Bumble, Peter Day, Dennis Harding, Jonathan Kawamura, Henry LeDuc, William McGrath, and Steve Monacos and Bertrand Delaet (CEA-LETI) for technical contributions at JPL, and Yale colleagues Faustin Carter and Luigi Frunzio, and Bertrand Reulet (Sherbrooke University) for technical contributions at Yale.

## REFERENCES

[1]    A. W. Blain, I. Smail, R. J. Ivison, J.-P. Kneib, and D. T. Frayer, "Submillimeter galaxies," *Physics Reports*, vol. 369, pp. 111–176, 2002.

[2]    J. Zmuidzinas and P. L. Richards, "Superconducting detectors and mixers for millimeter and submillimeter astrophysics," *Proceedings of the IEEE*, vol. 92, pp. 1597-1616, 2004.

[3]    C. M. Bradford and T. Nakagawa, "The future is BLISS - sensitive far-IR spectroscopy on SPICA and SAFIR," *New Astronomy Reviews*, vol. 50, pp. 221-227, 2006.

[4]    B. Swinyard, T. Nakagawa, and S. Consortium, "The space infrared telescope for cosmology and astrophysics: SPICA A joint mission between JAXA and ESA," *Experimental Astronomy*, vol. 23, pp. 193-219, Mar 2009.

[5]    D. Lester, J. Budinoff, and C. Lillie, "Large infrared telescopes in the exploration era: SAFIR," *Proceedings of SPIE*, vol. 6687, p. 66870M, 2007.

[6]    D. T. Leisawitz, T. Abel, R. J. Allen, D. J. Benford, A. Blain, C. Bombardelli, D. Calzetti, M. J. DiPirro, P. Ehrenfreund, N. J. E. II, J. Fischer, M. Harwit, T. T. Hyde, M. J. Kuchner, J. A. Leitner, E. C. Lorenzini, J. C. Mather, K. M. Menten, J. Samuel H. Moseley, L. G. Mundy, T. Nakagawa, D. A. Neufeld, J. C. Pearson, S. A. Rinehart, J. Roman, S. Satyapal, R. F. Silverberg, H. P. Stahl, M. R. Swain, T. D. Swanson, W. A. Traub, E. L. Wright, and H. W. Yorke, "SPECS: the kilometer-baseline far-IR interferometer in

NASA's space science roadmap," *Proceedings of SPIE*, vol. 5487, pp. 1527-1537, 2004.

[7]    P. F. Goldsmith, M. Bradford, M. Dragovan, C. Paine, C. Satter, B. Langer, H. Yorke, K. Huffenberger, D. Benford, and D. Lester, "CALISTO: the cryogenic aperture large infrared space telescope observatory," *Proceedings of SPIE*, vol. 7010, pp. 701020-16, 2008.

[8]    F. P. Helmich and R. J. Ivison, "FIRI—A far-infrared interferometer," *Experimenatl Astronomy*, vol. 23, pp. 245-276, 2009.

[9]    W. Wild, N. S. Kardashev, S. F. Likhachev, N. G. Babakin, V. Y. Arkhipov, I. S. Vinogradov, V. V. Andreyanov, S. D. Fedorchuk, N. V. Myshonkova, Y. A. Alexsandrov, I. D. Novokov, G. N. Goltsman, A. M. Cherepaschuk, B. M. Shustov, A. N. Vystavkin, V. P. Koshelets, V. F. Vdovin, T. de Graauw, F. Helmich, F. vander Tak, R. Shipman, A. Baryshev, J. R. Gao, P. Khosropanah, P. Roelfsema, P. Barthel, M. Spaans, M. Mendez, T. Klapwijk, F. Israel, M. Hogerheijde, P. vander Werf, J. Cernicharo, J. Martin-Pintado, P. Planesas, J. D. Gallego, G. Beaudin, J. M. Krieg, M. Gerin, L. Pagani, P. Saraceno, A. M. Di Giorgio, R. Cerulli, R. Orfei, L. Spinoglio, L. Piazzo, R. Liseau, V. Belitsky, S. Cherednichenko, A. Poglitsch, W. Raab, R. Guesten, B. Klein, J. Stutzki, N. Honingh, A. Benz, A. Murphy, N. Trappe, A. Raisanen, and C. Millimetron, "Millimetron-a large Russian-European submillimeter space observatory," *Experimental Astronomy*, vol. 23, pp. 221-244, Mar 2009.

[10]   M. Yun, J. W. Beeman, R. Bhatia, J. J. Bock, W. Holmes, L. Hustead, T. Koch, J. L. Mulder, A. E. Lange, A. D. Turner, and L. Wild, "Bolometric detectors for the Planck surveyor," *Proceedings of SPIE*, vol. 4855, pp. 136-147, 2003.

[11]   H. Takahashi, H. Shibai, M. Kawada, T. Hirao, T. Watabe, Y. Tsuduku, H. Nagata, H. Utsuno, Y. Hibi, S. Hirooka, T. Nakagawa, H. Kaneda, S. Matsuura, T. Kii, S. i. Makiuti, Y. Okamura, Y. Doi, H. Matsuo, N. Hiromoto, M. Fujiwara, and M. Noda, "FIS: far-infrared surveyor on board ASTRO-F (IRIS)," *Proceeding of SPIE*, vol. 4013, pp. 47-58, 2000.

[12]   S. E. Church, M. J. Griffin, M. C. Price, P. A. R. Ade, R. J. Emery, and B. M. Swinyard, "Performance characteristics of doped-Ge photoconductors for the Infrared Space Observatory Long Wavelength Spectrometer," *Proceeding of SPIE*, vol. 1946, pp. 116-125, 1993.

[13]   B. S. Karasik, W. R. McGrath, H. G. LeDuc, and M. E. Gershenson, "A hot-electron direct detector for radioastronomy," *Superconductor Science & Technology*, vol. 12, pp. 745-747, Nov 1999.

[14]   B. S. Karasik, W. R. McGrath, M. E. Gershenson, and A. V. Sergeev, "Photon-noise-limited direct detector based on disorder-controlled electron heating," *Journal of Applied Physics*, vol. 87, pp. 7586-7588, May 2000.

[15]   J. C. Mather, "Nobel lecture: From the big bang to the nobel prize and beyond," *Reviews of Modern Physics*, vol. 79, pp. 1331-1348, Oct-Dec 2007.

[16]   P. de Bernardis, P. A. R. Ade, J. J. Bock, J. R. Bond, J. Borrill, A. Boscaleri, K. Coble, B. P. Crill, G. De Gasperis, P. C. Farese, P. G. Ferreira, K. Ganga, M. Giacometti, E. Hivon, V. V. Hristov, A. Iacoangeli, A. H. Jaffe, A. E. Lange, L. Martinis, S. Masi, P. V. Mason, P. D. Mauskopf, A. Melchiorri, L. Miglio, T. Montroy, C. B. Netterfield, E. Pascale, F. Piacentini, D. Pogosyan, S. Prunet, S. Rao, G. Romeo, J. E. Ruhl, F. Scaramuzzi, D. Sforna, and N. Vittorio, "A flat Universe from high-resolution maps of the cosmic microwave background radiation," *Nature*, vol. 404, pp. 955-959, Apr 2000.

[17]   K. D. Irwin and G. C. Hilton, "Transition-edge sensors," in *Cryogenic Particle Detection*. vol. 99, ed Berlin: Springer-Verlag Berlin, 2005, pp. 63-149.

[18]   A. J. Miller, S. W. Nam, J. M. Martinis, and A. V. Sergienko, "Demonstration of a low-noise near-infrared photon counter with multiphoton discrimination," *Applied Physics Letters*, vol. 83, pp. 791-793, Jul 2003.

[19]   C. Groppi and J. H. Kawamura, "Coherent detector arrays for THz applications," *IEEE Transactions on Terahertz Science and Technology*, this issue.

[20]   A. A. Penzias and R. W. Wilson, "A measurement of excess antenna temperature at 4080 Mc/s," *Astrophysical Journal Letters*, vol. 142, pp. 419-421, Jul 1965.

[21] W. T. Reach, A. Abergel, F. Boulanger, F. X. Desert, M. Perault, J. P. Bernard, J. Blommaert, C. Cesarsky, D. Cesarsky, L. Metcalfe, J. L. Puget, F. Sibille, and L. Vigroux, "Mid-Infrared spectrum of the zodiacal light," *Astronomy and Astrophysics*, vol. 315, pp. L381-L384, Nov 1996.

[22] J. M. Lamarre, F. X. Desert, and T. Kirchner, "Background limited infrared and submillimeter instruments," *Space Science Reviews*, vol. 74, pp. 27-36, Oct 1995.

[23] C. Leinert, S. Bowyer, L. K. Haikala, M. S. Hanner, M. G. Hauser, A. C. Levasseur-Regourd, I. Mann, K. Mattila, W. T. Reach, W. Schlosser, H. J. Staude, G. N. Toller, J. L. Weiland, J. L. Weinberg, and A. N. Witt, "The 1997 reference of diffuse night sky brightness," *Astronomy & Astrophysics Supplement Series*, vol. 127, pp. 1-99, Jan 1998.

[24] J. M. Lamarre, "Photon noise in photometric instruments atfar-infrared and submillimeter wavelengths," *Applied Optics*, vol. 25, pp. 870-876, Mar 1986.

[25] P. L. Richards, "Bolometers for infrared and millimeter waves," *Journal of Applied Physics*, vol. 76, pp. 1-24, Jul 1994.

[26] E. H. Putley, "Indium antimonide submillimeter photoconductive detectors," *Applied Optics*, vol. 4, pp. 649-657, 1965.

[27] F. Arams, C. Allen, B. Peyton, and E. Sard, "Millimeter mixing and detection in bulk InSb," *Proceedings of the IEEE*, vol. 54, pp. 612-622, 1966.

[28] T. G. Phillips, "InSb heterodyne receivers for submillimeter astronomy," *Proceedings of SPIE*, vol. 280, pp. 101-107, 1981.

[29] E. M. Gershenzon, M. E. Gershenzon, G. N. Goltsman, A. D. Semenov, and A. V. Sergeev, "Heating of quasiparticles in a superconducting film in the resistive state," *JETP Letters*, vol. 34, pp. 268-270, 1981.

[30] E. M. Gershenzon, M. E. Gershenzon, G. N. Goltsman, A. D. Semenov, and A. V. Sergeev, "Electron heating in superconductor resistive state due to electromagnetic radiation," *Solid State Communications*, vol. 50, pp. 207-210, 1984.

[31] E. M. Gershenzon, M. E. Gershenson, G. N. Goltsman, A. D. Semenov, and A. V. Sergeev, "On the limiting characteristics of high-speed superconducting bolometers," *Zh. Tekh. Fiz.*, vol. 59, pp. 111-120, 1989; [*Sov. Phys. Tech. Phys.*, vol. 34, pp. 195-201, 1989].

[32] E. M. Gershenzon, G. N. Gol'tsman, I. G. Gogidze, Y. P. Gousev, A. I. Elant'ev, B. S. Karasik, and A. D. Semenov, "Millimeter and submillimeter range mixer based on electron heating of superconducting films in the resistive state," *Sverkhprovodimost (KIAE)*, vol. 3, pp. 2143-2160, 1990; [*Sov. Phys. Superconductivity*, vol. 3, pp. 1582-1597, 1990].

[33] E. M. Gershenzon, M. E. Gershenzon, G. N. Goltsman, B. S. Karasik, A. M. Lyul'kin, and A. D. Semenov, "Fast response superconducting electron bolometer," *Pis'ma Zh. Tekh. Fiz.*, vol. 15, pp. 88-92, Feb 1989; [*Sov. Tech. Phys. Lett.*, vol. 15, pp. 118-119, 1989].

[34] Y. P. Gousev, G. N. Goltsman, A. D. Semenov, E. M. Gershenzon, R. S. Nebosis, M. A. Heusinger, and K. F. Renk, "Broadband ultrafast superconducting NbN detector for electromagnetic radiaiton," *Journal of Applied Physics*, vol. 75, pp. 3695-3697, Apr 1994.

[35] Y. P. Gousev, G. N. Goltsman, B. S. Karasik, E. M. Gershenzon, A. D. Semenov, H. S. Barowski, R. S. Nebosis, and K. E. Renk, "Quasioptical superconducting hot electron bolometer for submillimeter waves," *International Journal of Infrared and Millimeter Waves*, vol. 17, pp. 317-331, Feb 1996.

[36] Y. P. Gousev, A. D. Semenov, G. N. Goltsman, A. V. Sergeev, and E. M. Gershenzon, "Electron-phonon interaction in disordered NbN films," *Physica B*, vol. 194, pp. 1355-1356, Feb 1994.

[37] B. S. Karasik, K. S. Ilin, E. V. Pechen, and S. I. Krasnosvobodtsev, "Diffusion cooling mechanism in a hot-electron NbC microbolometer mixer," *Applied Physics Letters*, vol. 68, pp. 2285-2287, Apr 1996.

[38] K. S. Il'in, N. G. Ptitsina, A. V. Sergeev, G. N. Gol'tsman, E. M. Gershenzon, B. S. Karasik, E. V. Pechen, and S. I. Krasnosvobodtsev, "Interrelation of resistivity and inelastic electron-phonon scattering rate in impure NbC films," *Physical Review B*, vol. 57, pp. 15623-15628, Jun 1998.

[39] D. E. Prober, "Superconducting terahertz mixer using a transition-edge microbolometer," *Applied Physics Letters*, vol. 62, pp. 2119-2121, Apr 1993.

[40] A. Skalare, W. R. McGrath, B. Bumble, H. G. LeDuc, P. J. Burke, A. A. Verheijen, R. J. Schoelkopf, and D. E. Prober, "Large bandwidth and low noise in a diffusion-cooled hot-electron bolometer mixer," *Applied Physics Letters*, vol. 68, pp. 1558-1560, Mar 1996.

[41] P. J. Burke, R. J. Schoelkopf, D. E. Prober, A. Skalare, B. S. Karasik, M. C. Gaidis, W. R. McGrath, B. Bumble, and H. G. LeDuc, "Mixing and noise in diffusion and phonon cooled superconducting hot-electron bolometers," *Journal of Applied Physics*, vol. 85, pp. 1644-1653, Feb 1999.

[42] E. M. Gershenzon, M. E. Gershenzon, G. N. Goltsman, A. D. Semenov, and A. V. Sergeev, "Wide-band high-speed Nb and YBaCuO detectors," *IEEE Transactions on Magnetics*, vol. 27, pp. 2836-2839, Mar 1991.

[43] M. A. Heusinger, A. D. Semenov, R. S. Nebosis, Y. P. Gousev, and K. F. Renk, "Nonthermal kinetic inductance photoresponse of thin superconducting films," *IEEE Transactions on Applied Superconductivity*, vol. 5, pp. 2595-2598, Jun 1995.

[44] M. Lindgren, V. Trifonov, M. Zorin, M. Danerud, D. Winkler, B. S. Karasik, G. N. Goltsman, and E. M. Gershenzon, "Transient resistive photoresponse of $YBa_2Cu_3O_{7-d}$ films using low-power 0.8 and 10.6 µm laser radiation," *Applied Physics Letters*, vol. 64, pp. 3036-3038, May 1994.

[45] M. Danerud, D. Winkler, M. Lindgren, M. Zorin, V. Trifonov, B. S. Karasik, G. N. Goltsman, and E. M. Gershenzon, "Nonequilibrium and bolometric photoresponse in patterned $YBa_2Cu_3O_{7-d}$ thin-films," *Journal of Applied Physics*, vol. 76, pp. 1902-1909, Aug 1994.

[46] V. A. Trifonov, B. S. Karasik, M. A. Zorin, G. N. Goltsman, E. M. Gershenzon, M. Lindgren, M. Danerud, and D. Winkler, "9.6 mu m wavelength mixing in a patterned $YBa_2Cu_3O_{7-d}$ thin film," *Applied Physics Letters*, vol. 68, pp. 1418-1420, Mar 1996.

[47] R. S. Nebosis, R. Steinke, P. T. Lang, W. Schatz, M. A. Heusinger, K. F. Renk, G. N. Goltsman, B. S. Karasik, A. D. Semenov, and E. M. Gershenzon, "Picosecond $YBa_2Cu_3O_{7-d}$ detector for far-infrared radiation," *Journal of Applied Physics*, vol. 72, pp. 5496-5499, Dec 1992.

[48] A. V. Sergeev, A. D. Semenov, P. Kouminov, V. Trifonov, I. G. Goghidze, B. S. Karasik, G. N. Goltsman, and E. M. Gershenzon, "Transparency of an $YBa_2Cu_3O_{7-d}$ film substrate interface for thermal phonons measured by means of voltage response to radiation," *Physical Review B*, vol. 49, pp. 9091-9096, Apr 1994.

[49] B. S. Karasik, W. R. McGrath, and M. C. Gaidis, "Analysis of a high-T-c hot-electron superconducting mixer for terahertz applications," *Journal of Applied Physics*, vol. 81, pp. 1581-1589, Feb 1997.

[50] M. Nahum and J. M. Martinis, "Novel hot-electron microbolometer," *Physica B*, vol. 194, pp. 109-110, Feb 1994.

[51] L. Kuzmin, D. Chouvaev, M. Tarasov, P. Sundquist, M. Willander, and T. Claeson, "On the concept of a normal metal hot-electron microbolometer for space applications," *IEEE Transactions on Applied Superconductivity*, vol. 9, pp. 3186-3189, Jun 1999.

[52] D. Chouvaev, L. Kuzmin, and M. Tarasov, "Normal-metal hot-electron microbolometer with on-chip protection by tunnel junctions," *Superconductor Science & Technology*, vol. 12, pp. 985-988, Nov 1999.

[53] L. S. Kuzmin, "On the concept of a hot-electron microbolometer with capacitive coupling to the antenna," *Physica B*, vol. 284, pp. 2129-2130, Jul 2000.

[54] A. T. Lee, P. L. Richards, S. W. Nam, B. Cabrera, and K. D. Irwin, "A superconducting bolometer with strong electrothermal feedback," *Applied Physics Letters*, vol. 69, pp. 1801-1803, 1996.

[55] B. Cabrera, R. M. Clarke, P. Colling, A. J. Miller, S. Nam, and R. W. Romani, "Detection of single infrared, optical, and ultraviolet photons using superconducting transition edge sensors," *Applied Physics Letters*, vol. 73, pp. 735-737, Aug 1998.

[56] A. E. Lita, A. J. Miller, and S. W. Nam, "Counting near-infrared single-photons with 95% efficiency," *Optics Express*, vol. 16, pp. 3032-3040, Mar 2008.

[57] D. Fukuda, R. M. T. Damayanthi, A. Yoshizawa, N. Zen, H. Takahashi, K. Amemiya, and M. Ohkubo, "Titanium based transition edge microcalorimeters for optical photon measurements," *IEEE Transactions on Applied Superconductivity*, vol. 17, pp. 259-262, Jun 2007.

[58]    D. Fukuda, G. Fujii, Y. A., T. H., R. M. T. Damayanthi, H. Takahashi, S. Inoue, and M. Ohkubo, "High speed photon number resolving detector with titanium transition edge sensor," *Journal of Low Temperature Physics,* vol. 151, pp. 100-105, 2008.

[59]    D. Bagliani, F. Gatti, M. Ribeiro Gomes, L. Parodi, L. Ferrari, and R. Valle, "Ir TES electron-phonon thermal conductance and single photon detection," *Journal of Low Temperature Physics,* vol. 151, pp. 234-238, 2008.

[60]    E. Taralli, C. Portesi, R. Rocci, M. Rajteri, and E. Monticone, "Investigation of Ti/Pd bilayer for single photon detection," *IEEE Transactions on Applied Superconductivity,* vol. 19, pp. 493-495, Jun 2009.

[61]    M. Rajteri, E. Taralli, C. Portesi, E. Monticone, and J. Beyer, "Photon-number discriminating superconducting transition-edge sensors," *Metrologia,* vol. 46, pp. S283-S287, 2009.

[62]    B. Cabrera and R. W. Romani, "Optical/UV astrophysics applications of cryogenic detectors," in *Cryogenic Particle Detection.* vol. 99, ed Berlin: Springer-Verlag Berlin, 2005, pp. 417-451.

[63]    *KISS Workshop 2010.* Available: http://www.kiss.caltech.edu/workshops/photon2009/index.html

[64]    A. D. Semenov, G. N. Gol'tsman, and R. Sobolewski, "Hot-electron effect in superconductors and its applications for radiation sensors," *Superconductor Science & Technology,* vol. 15, pp. R1-R16, 2002.

[65]    A. F. Andreev, "The thermal conductivity of the intermediate state in superconductors," *Soviet Physics JETP,* vol. 19, pp. 1228-1231, 1964.

[66]    M. O. Reese, "Superconducting hot electron bolometers for terahertz sensing," Ph.D., Applied Physics, Yale University, New Haven, 2006.

[67]    D. Golubev and L. Kuzmin, "Nonequilibrium theory of a hot-electron bolometer with normal metal-insulator-superconductor tunnel junction," *Journal of Applied Physics,* vol. 89, pp. 6464-6472, Jun 2001.

[68]    K. D. Irwin, "Phonon-mediated particle detection using superconducting tungsten trabsition-edge sensors," PhD thesis, Department of Physics, Stanford University, 1995.

[69]    K. D. Irwin, "An application of electrothermal feedback for high-resolution cryogenic particle-detection," *Applied Physics Letters,* vol. 66, pp. 1998-2000, Apr 1995.

[70]    P. Santhanam, S. Wind, and D. E. Prober, "Localization, superconducting fluctuations, and superconductivity in thin-films and narrow wires of aluminum," *Physical Review B,* vol. 35, pp. 3188-3206, Mar 1987.

[71]    J. Wei, D. Olaya, B. S. Karasik, S. V. Pereverzev, A. V. Sergeev, and M. E. Gershenson, "Ultrasensitive hot-electron nanobolometers for terahertz astrophysics," *Nature Nanotechnology,* vol. 3, pp. 496-500, Aug 2008.

[72]    B. S. Karasik, B. Delaet, W. R. McGrath, J. Wei, M. E. Gershenson, and A. V. Sergeev, "Experimental study of superconducting hot-electron sensors for submm astronomy," *IEEE Transactions on Applied Superconductivity,* vol. 13, pp. 188-191, Jun 2003.

[73]    B. S. Karasik and R. Cantor, "Optical NEP in hot-electron nanobolometers," in *21st International Symposium on Space Terahertz Technology,* 2010, pp. 291-297.

[74]    B. S. Karasik and R. Cantor, "Demonstration of high optical sensitiivty in far-infrared hot-electron bolometer," *Applied Physcs Letters,* vol. 98, p. 193503, May 2011.

[75]    J. Wei, "Hot-electron effects, energy transport and decoherence in nano-systmes at low temperatures," Ph.D. Ph.D. thesis, Department of Physics and Astronomy, Rutgers University, New Brunswick, 2007.

[76]    E. Taralli, M. Rajteri, E. Monticone, and C. Portesi, "Development of superconducting single-photon detector at I.N.Ri.M.," *International Journal of Quantum Information,* vol. 5, pp. 293-298, 2007.

[77]    M. E. Gershenson, D. Gong, T. Sato, B. S. Karasik, and A. V. Sergeev, "Millisecond electron-phonon relaxation in ultrathin disordered metal films at millikelvin temperatures," *Applied Physics Letters,* vol. 79, pp. 2049-2051, Sep 2001.

[78]    P. Day, H. G. Leduc, C. D. Dowell, R. A. Lee, A. Turner, and J. Zmuidzinas, "Distributed antenna-coupled TES for FIR detector arrays," *Journal of Low Temperature Physics,* vol. 151, pp. 477-482, Apr 2008.

[79]    A. Sergeev and V. Mitin, "Breakdown of Pippard ineffectiveness condition for phonon-electron scattering in micro and nanostructure," *Europhysics Letters,* vol. 51, p. 641, 2000.

[80]    A. V. Sergeev and V. Mitin, "Electron-phonon interaction in disordered conductors: static and vibrating scattering potentials," *Physical Review B,* vol. 61, p. 6041, 2000.

[81]    A. B. Pippard, "Ultrasonic attenuation in metals," *Philosophical Magazine,* vol. 46, p. 1104, 1955.

[82]    A. V. Sergeev and M. Y. Reizer, "Photoresponse mechanisms of thin superconducting films and superconducting detectors," *International Journal of Modern Physics B,* vol. 10, pp. 635-667, Mar 1996.

[83]    Y.-L. Zhong, A. Sergeev, C.-D. Chen, and J.-J. Lin, "Direct observation of electron dephasing due to inelastic scattering from defects in weakly disordered AuPd wires," *Physical Review Letters,* vol. 104, p. 206803, 2010.

[84]    A. Sergeev, B. S. Karasik, M. Gershenson, and V. Mitin, "Electron-phonon scattering in disordered metallic films," *Physica B,* vol. 316, pp. 328-330, May 2002.

[85]    B. S. Karasik, S. V. Pereverzev, D. Olaya, M. E. Gershenson, R. Cantor, J. H. Kawamura, P. K. Day, B. Bumble, H. G. LeDuc, S. P. Monacos, D. G. Harding, D. Santavicca, F. Carter, and D. E. Prober, "Development of the nano-HEB array for low-background far-IR applications," *Proceedings of SPIE,* vol. 7741, p. 774119, 2010.

[86]    J. Lin and J. P. Bird, "Recent experimental studies of electron dephasing in metal and semiconductor mesoscopic structures," *Journal of Physics: Condensed Matter,* vol. 14, p. R501, 2002.

[87]    E. M. Gershenzon, M. E. Gershenzon, G. N. Goltsman, A. Lulkin, A. D. Semenov, and A. V. Sergeev, "Electron-phonon interaction in ultrathin Nb films," *JETP,* vol. 70, 1990.

[88]    B. S. Karasik, S. V. Pereverzev, D. Olaya, J. Wei, M. E. Gershenson, and A. V. Sergeev, "Noise measurements in hot-electron titanium nanobolometers," *IEEE Transactions on Applied Superconductivity,* vol. 19, pp. 532-535, Jun 2009.

[89]    J. C. Mather, "Bolometer noise - non-equilibrium theory," *Applied Optics,* vol. 21, pp. 1125-1129, 1982.

[90]    B. S. Karasik and A. I. Elantiev, "Noise temperature limit of a superconducting hot-electron bolometer mixer," *Applied Physics Letters,* vol. 68, pp. 853-855, Feb 1996.

[91]    A. Sergeev, B. S. Karasik, N. G. Ptitsina, G. M. Chulkova, K. S. Il'in, and E. M. Gershenson, "Electron-phonon interaction in disordered conductors," *Physica B,* vol. 263, pp. 190-192, Mar 1999.

[92]    D. Morozov, P. D. Mauskopf, P. Ade, M. Ridder, P. Khosropanah, M. Bruijn, J. van der Kuur, H. Hoevers, J. R. Gao, and D. Griffin, "Ultrasensitive TES bolometers for space based FIR astronomy," *IEEE Transactions on Applied Superconductivity,* 2011, in press.

[93]    V. I. Chernichkin, D. E. Dolzhenko, L. I. Ryabova, L. V. Nicorici, and D. R. Khokhlov, "Sensitive direct detectors of terahertz radiation based on Pb1-xSnxTe(In)," *Journal of Infrared, Millimeter, and Terahertz Waves,* (2011), submitted.

[94]    C.-C. Chen, "Transmission of microwave through perforated flat plates of finite thickness," *IEEE Transactions on Microwave Theory and Techniques,* vol. 21, pp. 1-6, 1973.

[95]    B. S. Karasik and A. V. Sergeev, "THz hot-electron photon counter," *IEEE Transactions on Applied Superconductivity,* vol. 15, pp. 618-621, Jun 2005.

[96]    C. Beichman, P. Lawson, O. Lay, A. Ahmed, U. S., and K. Johnston, "Status of the terrestrial planet finder interferometer (TPF-I)," *Proceedings of SPIE,* vol. 6268, p. 62680S, 2006.

[97]    D. Des Marais, M. O. Harwit, K. W. Jucks, J. F. Kasting, D. N. C. Lin, J. I. Lunine, J. Schneider, S. Seager, W. A. Traub, and N. J. Woolf, "Remote sensing of planetary properties and biosignatures on extrasolar terrestrial planets," *Astrobiology,* vol. 2, pp. 153-181, 2002.

[98]    D. F. Santavicca, B. Reulet, B. S. Karasik, S. V. Pereverzev, D. Olaya, M. E. Gershenson, L. Frunzio, and D. E. Prober, "Energy resolution of terahertz single-photon-sensitive bolometric detectors," *Applied Physics Letters,* vol. 96, p. 083505, 2010.

[99]    C. M. Wilson and D. E. Prober, "Quasiparticle number fluctuations in superconductors," *Physical Review B,* vol. 69, Mar 2004.

[100] M. Kenyon, P. K. Day, C. M. Bradford, J. J. Bock, and H. G. Leduc, "Progress on background-limited membrane-isolated TES bolometers for far-IR/submillimeter spectroscopy," *Proceedings of SPIE*, vol. 6275, p. 627508, 2006.

[101] P. D. Mauskopf, P. A. R. Ade, J. Beyer, M. Bruijn, J. R. Gao, D. Glowaca, D. Goldie, D. Griffin, M. J. Griffin, F. C. Hoevers, P. Khosropanah, P. Kooijma, P. A. J. De Korte, D. Morozov, A. Murphy, C. O'Sullivan, M. Ridder, N. Trappe, H. Van Weers, J. Van Der Kuur, and S. Withington, "A TES focal plane for SPICA-SAFARI," in *21st International Symposium on Space Terahertz Technology*, 2010, pp. 246-255.

[102] P. Khosropanah, R. Hijmering, M. Ridder, L. Gottardi, M. Bruijn, J. v. d. Kuur, P. A. J. de Korte, J. R. Gao, F. C. Hoevers, D. Morozov , and P. Mauskopf, "Low noise transition edge sensor (TES) for SAFARI instrument on SPICA," *Proc. 22nd. Int. Symp. Space THz Technol., Tucson, AZ, 26-28 April, 2011,* in press.

[103] D. Morozov, P. Mauskopf, P. A. R. Ade , D. Griffin, J. R. Gao, F. C. Hoevers, P. Khosropanah, R. Hijmering, M. Ridder, and M. Bruijn, "Optical characterizaiton of high sensitivity TES detectors designed for the SPICA/SAFARI," *Proc. 22nd. Int. Symp. Space THz Technol., Tucson, AZ, 26-28 April, 2011,* in press.

[104] A. V. Sergeev, V. V. Mitin, and B. S. Karasik, "Ultrasensitive hot-electron kinetic-inductance detectors operating well below the superconducting transition," *Applied Physics Letters,* vol. 80, pp. 817-819, Feb 2002.

[105] P. K. Day, H. G. LeDuc, B. A. Mazin, A. Vayonakis, and J. Zmuidzinas, "A broadband superconducting detector suitable for use in large arrays," *Nature,* vol. 425, pp. 817-821, Oct 2003.

[106] A. Monfardini, L. J. Swenson, A. Bideaud, F. X. Desert, S. J. C. Yates, A. Benoit, A. M. Baryshev, J. J. A. Baselmans, S. Doyle, B. Klein, M. Roesch, C. Tucker, P. Ade, M. Calvo, P. Camus, C. Giordano, R. Guesten, C. Hoffmann, S. Leclercq, P. Mauskopf, and K. F. Schuster, "NIKA: A millimeter-wave kinetic inductance camera," *Astronomy & Astrophysics,* vol. 521, Oct 2010.

[107] H. G. Leduc, B. Bumble, P. K. Day, B. H. Eom, J. S. Gao, S. Golwala, B. A. Mazin, S. McHugh, A. Merrill, D. C. Moore, O. Noroozian, A. D. Turner, and J. Zmuidzinas, "Titanium nitride films for ultrasensitive microresonator detectors," *Applied Physics Letters,* vol. 97, Sep 2010.

[108] P. J. de Visser, J. J. A. Baselmans, P. Diener, S. J. C. Yates, A. Endo, and T. M. Klapwijk, "Number fluctuations of sparse quasiparticles in a superconductor," *Physical Review Letters,* vol. 106, p. 167004, Apr 2011.

[109] R. J. Schoelkopf, S. H. Moseley, C. M. Stahle, P. Wahlgren, and P. Delsing, "A concept for a submillimeter-wave single-photon counter," *IEEE Transactions on Applied Superconductivity,* vol. 9, pp. 2935-2939, 1999.

[110] D. E. Prober, J. D. Teufel, C. M. Wilson, L. Frunzio, M. Shen, R. J. Schoelkopf, T. R. Stevenson, and E. J. Wollack, "Ultrasensitive quantum-limited far-infrared STJ detectors," *IEEE Transactions on Applied Superconductivity,* vol. 17, pp. 241-245, 2007.

[111] M. D. Shaw, J. Bueno, P. Day, C. M. Bradford, and P. M. Echternach, "Quantum capacitance detector: A pair-breaking radiation detector based on the single Cooper-pair box," *Physical Review B,* vol. 79, p. 144511, Apr 2009.

[112] J. Bueno, M. D. Shaw, P. K. Day, and P. M. Echternach, "Proof of concept of the quantum capacitance detector," *Applied Physcs Letters,* vol. 96, p. 103503, 2010.

[113] S. Komiyama, O. Astafiev, V. Antonov, T. Kutsuwa, and H. Hirai, "A single-photon detector in the far-infrared range," *Nature,* vol. 403, pp. 405-407, Jan 2000.

[114] O. Astafiev, S. Komiyama, T. Kutsuwa, V. Antonov, Y. Kawaguchi, and K. Hirakawa, "Single-photon detector in the microwave range," *Applied Physics Letters,* vol. 80, pp. 4250-4252, Jun 2002.

[115] S. Komiyama, "Single-photon detectors in the terahertz range," *IEEE Journal of Selected Topics in Quantum Electronics,* vol. 17, pp. 54-66, Jan-Feb 2011.

[116] H. Hashiba, V. Antonov, L. Kulik, A. Tzalenchuk, and S. Komiyama, "Sensing individual terahertz photons," *Nanotechnology,* vol. 21, Apr 2010.

[117] K. Ikushima, Y. Yoshimura, T. Hasegawa, S. Komiyama, T. Ueda, and K. Hirakawa, "Photon-counting microscopy of terahertz radiation," *Applied Physics Letters,* vol. 88, Apr 2006.

[118] B. A. Young, J. R. Williams, S. W. Deiker, S. T. Ruggiero, and B. Cabrera, "Using ion implantation to adjust the transition temperature of superconducting films," *Nuclear Instruments & Methods in Physics Research A,* vol. 520, pp. 307-310, 2004.

[119] B. A. Young, T. Saab, B. Cabrera, J. J. Cross, R. M. Clarke, and R. A. Abusaidi, "Measurement of Tc suppression in tungsten using magnetic impurities," *Journal of Applied Physics,* vol. 86, pp. 6975-6978, Dec 1999.

[120] S. W. Deiker, W. Doriese, G. C. Hilton, K. D. Irwin, W. H. Rippard, J. N. Ullom, R. L. Vale, S. T. Ruggiero, A. Williams, and B. A. Young, "Superconducting transition edge sensor using dilute AlMn," *Applied Physcs Letters,* vol. 85, pp. 2137-2139, 2004.

[121] B. S. Karasik, P. K. Day, J. H. Kawamura, B. Bumble, and H. G. LeDuc, "Multiplexing of Hot-Electron Nanobolometers Using Microwave SQUIDs," *AIP Conference Proceedings,* vol. 1185, pp. 257-260, 2009.

[122] W. Zhang, P. Khosropanah, J. R. Gao, T. Bansal, T. M. Klapwijk, W. Miao, and S. C. Shi, "Noise temperature and beam pattern of an NbN hot electron bolometer mixer at 5.25 THz," *Journal of Applied Physics,* vol. 108, p. 093102, Nov 2010.

[123] A. D. Semenov, H. Richter, H. W. Hubers, B. Gunther, A. Smirnov, K. S. Il'in, M. Siegel, and J. P. Karamarkovic, "Terahertz performance of integrated lens antennas with a hot-electron bolometer," *IEEE Transactions on Microwave Theory and Techniques,* vol. 55, pp. 239-247, Feb 2007.

[124] E. N. Grossman, J. E. Sauvageau, and D. G. McDonald, "Lithographic spiral antennas at short wavelengths," *Applied Physics Letters,* vol. 59, pp. 3225-3227, Dec 1991.

[125] J. Alda, J. M. Rico-García, J. M. López-Alonso, and G. Boreman, "Optical antennas for nano-photonic applications," *Nanotechnology,* vol. 16, pp. S230-S234, 2005.

[126] F. J. Gonzalez, B. Ilic, J. Alda, and G. D. Boreman, "Antenna-coupled infrared detectors for imaging applications," *IEEE Journal of Selected Topics in Quantum Electronics,* vol. 11, pp. 117-120, Jan-Feb 2005.

[127] D. R. Schmidt, R. J. Schoelkopf, and A. N. Cleland, "Photon-mediated thermal relaxation of electrons in nanostructures," *Physical Review Letters,* vol. 93, p. 045901, Jul 2004.

[128] G. Temporao, H. Zbinden, S. Tanzilli, N. Gisin, T. Aellen, M. Giovanni, J. Faist, and J. P. von der Wied, "Feasibility study of free-space qunatum key distribution in the mid-infrared," *Quantum Information and Computation,* vol. 8, pp. 1-11, 2008.

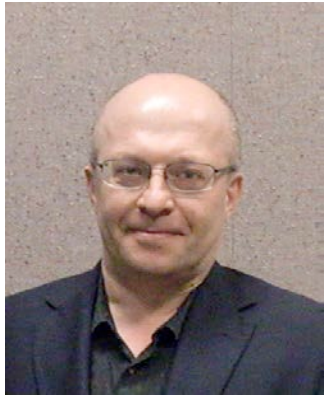

**Boris S. Karasik** received his BS degree in physics from Petrozavodsk State University (1979) and a PhD in physics from Moscow State Pedagogical University (1988).

He participated in many Russian, US, and European projects on hot-electron device physics and applications. He joined Jet Propulsion Laboratory (JPL), California Institute of Technology, in 1995 as a Member of Technical Staff and is currently a Principal Research Technologist. He has some 135 journal and conference publications and 6 patents. He is a recipient of several NASA and JPL awards and certificates of recognition for technical achievements. He is a member of the American Physical Society and of Sigma Xi.

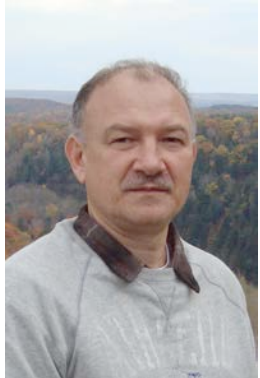

**Andrei V. Sergeev** received his BS and MS degrees in physics from Moscow Institute of Physics and Technology (MIPT) in 1978 and 1980, respectively. He received Ph.D. from Moscow State Pedagogical University in theoretical solid state physics in 1987.

His research interests are concentrated in the areas of quantum kinetics and transport in nanostructures and modeling of nanodevices and nanosensors. Specific research topics include hot-electron nano-detectors, single-photon counters, nanocalorimeters, wide-band mixers, quantum-dot detectors, solar cells. His design of advanced devices is based on multi-scale modeling, which includes the quantum transport, hydrodynamic transport, and Monte-Carlo simulations. He has published 95 scientific papers in refereed journals. Presently, he is with the University at Buffalo SUNY.

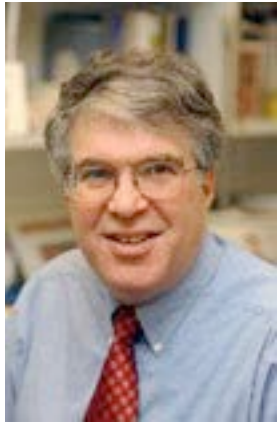

**Daniel E. Prober** received a BS in physics from Brandeis University, Waltham, MA in 1970 and MS (1971) and PhD (1975) in physics from Harvard University, Cambridge, MA.

He has been a faculty member at Yale University, New Haven, CT since 1975, as Assistant Professor progressing through Professor of Applied Physics and Physics in 1986, and department chair of Applied Physics, 2003-9. He has received two Fulbright Fellowships, and the NASA Technical Innovation Award with colleagues at JPL for the Superconducting Hot-electron Bolometer. He is a Fellow of the American Physical Society.